\documentclass[preprint,prb]{revtex4}

\usepackage{graphicx}
\usepackage{dcolumn}
\usepackage{bm}
\usepackage{amsmath}
\usepackage{amssymb}
\usepackage{setspace}
\usepackage{caption}
\usepackage{color}
\begin{document}


\title{Intermolecular Forces and the Glass Transition}

\author{Randall W. Hall\footnote{Corresponding author, rhall@lsu.edu}}
\affiliation{Department of Chemistry, Louisiana State University\\
Baton Rouge, LA, 70808}
\author{Peter G. Wolynes}%
\affiliation{Department of Chemistry and Biochemistry\\
University of California\\ San Diego, CA }

\date{\today}

\begin{abstract}
Random first order transition theory is used to determine the role of attractive and repulsive interactions in the dynamics of supercooled liquids.  Self-consistent phonon theory, an approximate mean field treatment consistent with random first order transition theory, is used to treat individual glassy configurations, while the liquid phase is treated using common liquid state approximations.  Free energies are calculated using liquid state perturbation theory.  The  transition temperature $T^{*}_{A} $, the temperature where the onset of activated behavior is predicted by mean field theory, the lower crossover temperature $T_{c}^{*}$ where barrierless motions actually occur through fractal or stringy motions (corresponding to the phenomenological mode coupling transition temperature), and $T^{*}_{K} $, the Kauzmann temperature (corresponding to an extrapolated entropy crisis), are calculated in addition to $T^{*}_{g} $, the glass transition temperature that corresponds to laboratory cooling rates.  Relationships between these quantities agree well with existing experimental and simulation data on van der Waals liquids.  Both the isobaric and isochoric behavior in the  supercooled regime are studied, providing results for  $\Delta C_{V} $ and $\Delta C_{p} $ that can be used to calculate the fragility as a function of density and pressure, respectively.  The predicted variations in the $\alpha$-relaxation time with temperature and density conform to the empirical density-temperature scaling relations found by  Casalini and Roland. We thereby demonstrate the microscopic origin of their observations. Finally, the relationship first suggested by Sastry between the spinodal temperature and the Kauzmann temperatures, as a function of density, is examined. The present microscopic calculations support the existence of an intersection of these two temperatures at sufficiently low temperatures.
\end{abstract}

\pacs{61.43.Fs,64.70.Pf,65.60.+a}
\maketitle
\section{Introduction}

Owing to their mutual repulsion, molecules in a dense supercooled liquid can only rearrange by performing an intricately correlated dance. The growing complexity of the cooperative motion as the liquid cools leads to the glass transition. In the deeply supercooled state, a sufficiently large number of molecules move in the cooperative activated events so that the dynamics can be predicted rather accurately from the mesoscale thermodynamics of the configurations. The microscopic theory of glasses based on random first order transitions quantitatively predicts\cite{LubchenkoWolynes2007} the typical relaxation time\cite{Wolynes2000}, its distribution\cite{Wolynes2001A} and the size and geometry\cite{SchmalianWolynes2006} of the independently rearranging regions using only the knowledge of the configurational entropy which can be obtained to reasonable accuracy from experiment\cite{Speedy2001,Johari2002,Angell2002b,CommentOnExpt}.

Furthermore the random first order transition theory provides a route to calculate the relevant mesoscale configurational thermodynamics directly from the intermolecular forces. The theory thus allows one to determine what features of the intermolecular forces are important for the configurational thermodynamics and therefore what determines microscopically the glassy dynamics. These topics are the focus of this paper. 

Not unexpectedly the details of the repulsive forces are key. For glass physics a purely steric picture envisioning infinitely hard repulsion provides a good place to start, but, as we shall see, a truly quantitative treatment must go further. Since the time of Rice\cite{Rice1944} and Kirkwood\cite{Kirkwood1935} we have become used to the dominance of hardcore sterics in determining gross liquid structure. Alder and Wainwright's computer calculations\cite{Alder1962} and Widom's\cite{Widom1964} analytical treatment of freezing have made Newton's conjecture on the steric origin of crystals\cite{Opticks}  a quantitatively useful paradigm. Likewise the random first order transition theory shows the configurational entropy density of a hard sphere system will fall with decreasing volume leading to slowing of the dynamics and eventually results in a glass transition.

The impenetrable hard sphere picture, we shall see, however, is not quite accurate enough to describe real molecular liquids in the supercooled regime. High-pressure experiments make this failing clear. Apart from a trivial multiplicative temperature scaling of the time scale, the dynamics of hard spheres must only depend on density. In the supercritical and ordinary temperature regime above the melting point studies of transport show that diffusion and rotational diffusion do indeed depend primarily on density for van der Waal's liquids.  At constant volume the temperature dependence of transport coefficients is almost negligible\cite{Jonas1984}. Such a weak temperature dependence at constant volume  is not found in the glassy regime, in contrast. While the apparent activation energy for flow at the glass transition is smaller at constant volume than along lines of constant pressure, it is not negligible: the ratio of the constant volume and constant pressure activation energies is measured\cite{CasaliniRoland2005} to be about 1/2. Developing an empirical scaling of transport in this regime has attracted attention, most notably in a series of works by Casalini and Roland\cite{CR2004,CasaliniRoland2005, Casalini2005, CasaliniRoland2006,CasaliniRoland2006B}. The shape of the repulsive and attractive forces clearly must matter in the supercooled regime.

We will show how the random first order transition theory clarifies the origin of this isochoric temperature dependence of the relaxation rate. First it helps to recall that the validity of an essentially purely steric picture of gross liquid structure and of nonactivated transport under usual thermodynamic conditions can be explained by the existence of a small parameter: the ratio of a steepness length of the potential to the particle size itself. Arguments from liquid state perturbation theory show structural quantities on the scale of the particle diameter can be obtained by a so-called "blip function" expansion\cite{WCA1,WCA2}. The blip function, measuring the deviation between the model impenetrable particles and real ones, is nonzero only near the turning point of molecular collisions and covers the range of a steepness length. The steepness length measures the range of distances a particle can penetrate with a typical thermal kinetic energy, and is the order of 1/10 the particle diameter. Since the steepness length is small on the particle size scale, treating the particles as impenetrable becomes a good approximation. Random first order transition theory shows that another small parameter involving lengths emerges in the supercooled regime: the Lindemann ratio characterizing the extent of the local caged motions.

In the supercooled regime, molecules spend most of their time vibrating about fiducial locations characterizing the aperiodic free energy maxima. These vibrations also cover a length about 1/10 of the particle spacing revealed by both neutron scattering experiments\cite{Russina1999} and theoretical calculations\cite{Wolynes1984,Wolynes1985}.

The smallness of the Lindemann ratio ultimately explains the quantitative successes of random first order transition theory (which is an expansion about mean field theory) in predicting the universal patterns of fragility in structural glasses and supercooled liquids\cite{EastwoodWolynes2002}. The ratio between the Lindemann length and the steepness length is clearly, however, not particularly small. Because of this fact, the configurational entropy and the configurational heat capacity, which enter the supercooled dynamics, depend on volume and temperature jointly. This joint dependence explains the significant size of the isochoric activation energy for relaxation and transport.

To make these ideas clearer and more explicit in this paper we present detailed calculations of the configurational thermodynamics of the supercooled Lennard-Jones fluid using the tools provided by the random first order transition theory. These calculations are based on density functional\cite{Wolynes1985,Valls1999, Kawasaki2002} and self-consistent phonon approaches\cite{Wolynes1984, Fixman1969}  which can be justified by elegant replica methodology\cite{Sompolinksky1985,Wolynes1987, Crisanti1992,Monasson1995,Franz1995,Parisi1997,Mezard1999, Mezard1999b}. 
Related theories have also recently appeared.\cite{Schweizer2007,Schweizer2003,Archer2007,Archer2006}
In addition our calculations rely on reasonably clear, but somewhat uncontrolled, liquid state approximations like those commonly employed for liquids that are not particularly dense or highly supercooled. It will turn out these approximations reproduce quite well the laboratory observations on the pressure dependence of the glass transition of molecular liquids. While these approximations predict an ideal glass transition at low temperature and high density, in keeping with extrapolated data they cannot address the strict existence of an entropy crisis with controlled rigor. Fortunately taking an agnostic attitude towards that ideal transition does not impair the application of the present scheme to the currently accessible experimental regime.  It appears to be impossible to reach the long time scales where corrections to the mean field configurational entropy should appear\cite{EastwoodWolynes2002}. At the lowest density studied, the entropy crisis occurs in a temperature-density range that is close to the spinodal temperature for a liquid-gas transition.  As suggested by Sastry\cite{Sastry2000}, there is the possibility of a temperature and density that is both the spinodal temperature-density and the entropy crisis temperature and density. We investigate this possibility using our theory.

The organization of the paper is as follows.  We first describe the liquid state approximations we use to describe the overall thermodynamics of the dense liquid states.  We then discuss how the self-consistent phonon calculations for a single aperiodic free energy minimum can be carried out for smooth repulsive forces. The configurational thermodynamic properties follow by subtraction. We then present the resulting predictions for the Kauzmann temperature, laboratory glass transition temperature, mean field dynamical transition temperature, and the laboratory crossover transition temperature as functions of density.  In the following section we compare the predictions starting from the intermolecular forces with the patterns of laboratory experiments. We show how the theory reproduces the known Casalini-Roland scaling of the dynamics over the measured range but suggests the possibility of observing deviations from their scaling if a broader thermodynamic range can be probed. The theory also correctly predicts the ratio of isochoric and isobaric activation energies at the glass transition. We summarize our results in the conclusion.  

\section{Theory}
\subsection*{Microscopic Theories of the Glass and Liquid}
The universal features of the dynamics of supercooled liquids and glasses are explained by the random first order transition theory~\cite{Wolynes1987,Wolynes1989,Wolynes2000,Wolynes2001A}.  At the  mean field theory level the underlying microscopic framework provides a description of several characteristic transition temperatures that are expected in glassy systems.  This theory predicts there is a temperature, $T^{*}_{A}$, at which a dynamic transition occurs, corresponding to the onset of activated motion in a liquid. At the mean field level, the mode coupling theory temperature $T^{*}_{MCT}$ at which the density-density correlation function no longer decays\cite{Wolynes1987} and coincides with  $T^{*}_{A} $. We note, however, that phenomenological fits to mode coupling theory consistently give a lower temperature than the ab initio mean field predictions. At the mean field level there is also a temperature, $T^{*}_{K}$, which corresponds to an entropy crisis predicted long ago by Kauzmann\cite{Kauzmann1948}.  Random first order transition theory goes beyond mean field theories to analyze the creation of so-called ``entropic droplets" which describe the nucleation  of small regions containing a multiplicity of states in an aperiodic free energy minimum. These droplets arise as corrections to the mean field  with a large ensemble of local minima of the mean field free energy functional for the  inside of the droplet in an initial mean field solution. Their multiplicity corresponds to the ``configurational entropy".  The size of the critically rearranging droplet is thus determined by a balance of the configurational entropy and the free energy penalty required to create a surface between two distinct aperiodic structures.  
	
Dynamic heterogeneities consistent with the mosaic structures arising from the entropic droplets predicted by random first order transition theory nearly twenty years ago have recently been observed directly in supercooled liquids\cite{Spiess1998, Israeloff2000, Cicerone1996}.  More recently, random first order transition calculations have quantitatively predicted the magnitude of the  barriers to reorganization of a mosaic region, as well as the fluctuations in barrier height from one region to another with using adjustable parameters\cite{Wolynes2000}.  The theory has also been applied to the diffusion of probe molecules in \textit{o}-terphenyl\cite{Xia2001} and to aging in structural glasses\cite{Wolynes2003}, again in good agreement with observation.  Of particular relevance to our work is the microscopic prediction of a typical relaxation time in the equilibrium supercooled liquid
\begin{eqnarray}
\tau &=&\tau_{0} \exp\left(\frac{\Delta F^{\ddagger}}{k_{B} T}\right)\label{Sceqn}\\
\frac{\Delta F^{\ddagger}}{k_{B} T} &=&\frac{32}{S_{c} / N k_{B} }\\
&\approx&\frac{32 T_{K}}{\left(\Delta C_{p} / N k_{B} \right) \left(T-T_{K} \right)}\\
&\equiv&\frac{D\ T_{K} }{T-T_{K} },
\end{eqnarray}
We see the predictions conform to the well-known empirical relation known as the Vogel-Tamann-Fucher law for the $\alpha$-relaxation time $\tau$.  In this expression,  $S_{c} $ is the configurational entropy per spherical unit, called a "bead."  At constant pressure, its temperature derivative is proportional to  $\Delta C_{p} $, the heat capacity jump at constant pressure. $D$ is often called the liquid's fragility. Eqn.~\ref{Sceqn} can be used to relate the the configurational entropy to the relaxation time scale in the liquid:
\begin{eqnarray}
\frac{S_{c} }{N k_{B} }&=&\frac{32}{\ln\tau/\tau_{0} }\label{timescale}
\end{eqnarray}
For the one hour time scale we see the configurational entropy per bead is predicted to be a universal value, $S_{c} / N k_{B} $ = 0.82.  The prediction that $D = 32/\left(\Delta C_{p} / N k_{B} \right)$ conforms very well to  the experimental data  for a variety of glasses including both molecular and ionic glasses.  To make this fit quantitative, one uses the melting entropy to count beads, as described by Lubchenko and Wolynes\cite{Wolynes2003} and by Stevenson and Wolynes\cite{Stevenson2005}. A survey of experimental data reveals that the predicted microscopic coefficient in this expression of 32 is only in error by roughly 10\%.

A convenient mean field treatment  of the localized phase of a random first order transition is provided by  self-consistent phonon theory, which envisions the atoms in a glass vibrating about fiducial sites in an effective potential due to the other atoms.  The effective vibrational frequency is calculated self-consistently as a function of density.  For the hard sphere glass there exists a mean field dynamical transition density $\rho_{A} $ below which the only solution to the self-consistent equations is a vibrational frequency of zero, corresponding to a mobile liquid where the dynamics are essentially those of a renormalized gas; above this density a new, non-zero, frequency appears.  For short times, the system is therefore dynamically a solid and corresponds to a glassy local minimum in the free energy.  The difference $F_{glass}-F_{liquid}\equiv T\ S_{c} $ is accounted for by  the configurational entropy of configurations.   At a high enough density the entropy is predicted by the mean field theory to no longer to be extensive.  The limit $S_{c} = 0$ defines $\rho_{K} $. The glass density $\rho_{g} $ can be calculated but depends on  the experimental time scale of interest (see below). For systems with potential energy functions other than purely hard sphere, there are corresponding temperatures $T_{A}$, $T_{g}$, and $T_{K}$.
While the mean field transition to rigidity occurs at $T_{A} $, fractal or string-like excitations still provide a downhill escape from minima at that temperature.  Only  at a  crossover temperature $T_{c}^{*}$ lower than the mean field $T^{*}_{A}$ do such string motions become activated thus explaining the overestimation of $T^{*}_{c}$ by ab initio mode coupling. Indeed, such a crossover is found where phenomenological fits to mode coupling theory show a transition\cite{Fayer2000} and deviations from Vogel-Tamman-Fulcher behavior are seen in a Stickel plot\cite{Stickel1996}. Thus, microscopic calculations based on  random first order transition theory predict\cite{SchmalianWolynes2006} a crossover temperature $T_{c} $ at a critical entropy
\begin{eqnarray}
S_{c}(T_{c} )/ N k_{B} &=& 1.28
\end{eqnarray}
$T_{c}$ is the temperature at which a complete transition from collisional to activated transport will be noticed. As a consequence of the universality of the crossover entropy, the relaxation time at the crossover is predicted to be universal, as indeed is observed\cite{SchmalianWolynes2006}.

Originally developed by  
Fixman to study periodic crystalline solids\cite{Fixman1969}, self-consistent phonon theory was used long ago by Stoessel  and Wolynes\cite{Wolynes1984} to calculate  $\rho_{A} $ in a monatomic hard sphere glass.  More recently, Hall and Wolynes\cite{HallWolynes2003} used this approach to study a model of a  network glass in which  the are both steric repulsions and athermal bonding constraints. Those calculations predict the well-known decrease in fragility with increasing connectivity empirically observed in mixtures of network formers.  We see that the mean field random first order transition theory based on self-consistent phonon ideas is not restricted to hard sphere potential functions and can therefore be applied directly to the Lennard-Jones glass.

Microscopic treatments of liquids traditionally take the view that because the structure of the liquid is dominated by repulsive forces, a good approach is to develop  perturbation theories based on a hard sphere reference system. The main issue is to find the optimal parameters for the reference system.  At densities less than the glass transition density, such approaches provide reasonably accurate overall thermodynamics but  at the higher densities accompanying the glass transition, less has been explored  about the accuracy of the different perturbation approaches.  In light of the similarity between the "blip" and Lindemann length scales because we lack  a non-perturbative approach to the Lennard-Jones liquid, it is  necessary to  examine two related but distinct detailed perturbation theories to treat the liquid state thermodynamics.  For consistency, we also employ  a perturbative self-consistent phonon theory. We now discuss the perturbation theories for liquid and glass.

\subsection*{Liquid State Thermodynamics and the Potential}
We carried out the calculations using two perturbation theories for the free energy of the Lennard-Jones liquid that relate the properties of the Lennard-Jones system to the purely steric hard sphere system. The first perturbation approach separates the potential along the lines of  the well-known Weeks-Chandler-Andersen (WCA) approximation\cite{WCA1,WCA2}.  This approximation is often used to treat molecular liquids much above the melting point.  A second  approximation, known to be more accurate for high density fluids, was developed by Kang, Ree, and Ree (KRR)\cite{KRR1,KRR2}.  In both schemes, the pair-potential for the system is written as a sum of an attractive and a repulsive term:
\begin{eqnarray}
V(r)&=&v(r) + w(r)\\
&\approx&v_{HS}(r)+w(r)
\end{eqnarray}
where $v_{HS}$ is a hard-sphere potential whose diameter $d$ is determined by
\begin{eqnarray}
\int d\mathbf{r}\left[\exp(-v(r)/k_{B} T)-\exp(-v_{HS}/k_{B} T)\right]y_{HS}(r)&=&0
\label{blip}
\end{eqnarray}
With either separation the Helmholtz free energy is obtained by adding a first order perturbation to the free energy of the corresponding hard sphere system:
\begin{eqnarray}
A&\approx&A_{HS}+\frac{N\rho}{2}\int d\mathbf{r}g(r)w(r)
\end{eqnarray}
The pair correlation can also be obtained:
\begin{eqnarray}
g(r)&\approx&y_{HS}(r)\exp(-v(r)/k_{B} T)\label{g_of_r}
\end{eqnarray}
where $y_{HS}$ is the cavity distribution.  
The pressure can be obtained by numerical differentiating the free energy: 
\begin{eqnarray}
p&=&-\left(\frac{\partial A}{\partial V}\right)_{T,N}
\end{eqnarray}
The specific separation of the Lennard-Jones potential
\begin{eqnarray}
V(r) &=& 4\epsilon \left[\left(\frac{\sigma}{r}\right)^{12}-\left(\frac{\sigma}{r}\right)^{6}\right]
\end{eqnarray}
due to WCA breaks the potential at its minimum
\begin{eqnarray}
\begin{tabular}{lllll}
$v(r)$&=&$V(r)+\epsilon$&&$r<2^{1/6}\sigma$\\
&=&0&&$r\ge 2^{1/6}\sigma$\\
$w(r)$&=&$-\epsilon$&&$r<2^{1/6}\sigma$\\
&=&$V(r)$&&$r\ge 2^{1/6}\sigma$
\end{tabular}
\label{KA}
\end{eqnarray}
The KRR separation uses an optimized but density dependent cutoff for separating the potential 
\begin{eqnarray}
\begin{tabular}{llll}
$v(r)$&=&$V(r)-F(r)$&$r\le \lambda$\\
&=&0&$r> \lambda$\\
$w(r)$&=&$F(r)$&$r \le \lambda$\\
&=&$V(r)$&$r > \lambda$
\end{tabular}
\label{KRR}
\end{eqnarray}
\begin{eqnarray}
F(r)&=&V(\lambda) - V^{'}(\lambda)\left(\lambda -r\right)
\end{eqnarray}
\begin{eqnarray}
\begin{tabular}{llll}
&=&$r_{m} $&$\rho < \rho_{1} $\\
$\lambda(\rho)$&=&$r_{m} +\frac{a^{'}(\rho_{2} )\left(\rho-\rho_{1} \right)^{3} \left[k\left(4\rho_{2} - \rho_{1} -3\rho\right)+3(\rho-\rho_{1} )\right]}{12\left(\rho_{2} -\rho_{1} \right)^{3} }$&$\rho_{1} \le \rho \le \rho_{2} $\\
&=&$a(\rho)$&$\rho_{2} < \rho$
\end{tabular}
\label{srho}
\end{eqnarray}
where $r_{m} = 2^{1/6}\sigma$, $a(\rho) = 2^{1/6}/\rho^{1/3}$, $a^{'}(\rho_{2} ) = (\partial a(\rho)/\partial \rho)_{\rho=\rho_{2} }$, $k = 3.026462$, $\rho_{1} = 0.99\rho_{c} $, $\rho_{2} = 1.01\rho_{c} $, and $\rho_{c} \sigma^{3} = \sqrt{2}/r_{m}^{3}$.   By design, the KRR separation reduces to the WCA separation at sufficiently low density.
For the remainder of this paper, we use units where $\sigma =1 $. 

For liquids above the melting point, $g_{HS}$ and $y_{HS}$ are well described by the Percus-Yevick approximation.  For the high densities considered in this work, it is necessary to employ  a more accurate modification of these pair functions similar to that made by Verlet and Weis. The Verlet-Weis correction improves the behavior of g(r) near contact and removes oscillations in g(r) for large r.  We follow the procedures of Robles and L\'{o}pez de Haro\cite{LopezdeHaro} and Verlet and Weis\cite{VW}.
First, for $r\ge d$, we write $g(r)$ as
\begin{eqnarray}
g_{HS}(r,d,\eta)&=&g_{PY}(r/d',\eta')+\delta g(r)\\
&=&g_{PY}(r/d',\eta')+\frac{B}{r}\exp(-\mu (r-d))\cos(\mu (r-d))
\end{eqnarray}
where $\eta=\rho^{*} d^{3} / 6$, $\eta' = \eta-\eta^{2} /16$, $\eta' = \rho^{*} d^{'3}$,  $g_{PY}$ is the Percus-Yevick hard-sphere g(r), and $B$ and $\mu$ are to be determined. $g(r)$ has structure extending for several $\sigma$ at the low temperatures and high densities studied in this work.  Therefore, we used the analytical form of Smith and Henderson\cite{Henderson1970} to generate values for $g_{PY}(r)$ out to 8 $\sigma$.

For $r<d$, we follow WCA\cite{WCA2} and extrapolate $\ln\delta g(r)$ quadratically about $r=d$:
\begin{eqnarray}
\ln\delta g_{<}(r)&=&\ln\frac{B}{d} -\left(\frac{1}{d}+\mu\right)(r-d)+\frac{1}{2}\left(\frac{1}{d^{2} }-\mu^{2} \right)(r-d)^{2} 
\end{eqnarray}
leading to
\begin{eqnarray}
y_{HS}(r,d,\eta)&=&y_{PY}(r/d',\eta')+\delta g_{<}(r)
\label{y_of_r}
\end{eqnarray}
 Eqn.~\ref{y_of_r} is used in Eqn.~\ref{blip} to determine the hard-sphere diameter $d$.

$B$ is determined by the value of $g(r)$ at contact, through the pressure equation:
\begin{eqnarray}
\beta p/\rho\equiv Z&=&1+4\eta g_{HS}(d,d,\eta)\\
&=&1+4\eta\left(g_{PY}(d/d',\eta')+\frac{B}{d}\right)
\end{eqnarray}
where $Z$ is the compressibility factor, which is determined given an equation of state.  Thus, given an equation of state $B$ can be found.  $\mu$ is determined by requiring consistency with the compressibility equation:
\begin{eqnarray}
k_{B} T\left(\frac{\partial p}{\partial \rho}\right)_{T,N}^{-1}&=&1+\rho\int d\mathbf{r} \left(g_{HS}(r,d,\eta)-1\right)\\
&=&\left(\frac{\partial \eta Z}{\partial \eta}\right)_{T,N}^{-1}\equiv \chi(Z)
\end{eqnarray}
or
\begin{eqnarray}
\chi(Z)&=&1+4\pi\rho\int_{d}^{\infty}dr\ r^{2} \left(g_{HS}(r,d,\eta)-1\right)\\
&=&1+4\pi\rho\int_{d}^{\infty}dr\ r^{2} \left(g_{PY}(r/d',\eta')-1\right)+4\pi\int_{d}^{\infty}dr\ r^{2} \delta g(r)\\
&=&1+4\pi\rho\int_{d'}^{\infty}dr\ r^{2} \left(g_{PY}(r/d',\eta')-1\right)+4\pi\rho\int_{d}^{d'}dr\ r^{2} \left(g_{PY}(r/d',\eta')-1\right)\nonumber \\
&&+4\pi\int_{d}^{\infty}dr\ r^{2} \delta g(r)\\
&=&\chi_{PY}+4\pi\rho\int_{d}^{d'}dr\ r^{2} \left(g_{PY}(r/d',\eta')-1\right)+4\pi\int_{d}^{\infty}dr\ r^{2} \delta g(r)
\end{eqnarray}
with
\begin{eqnarray}
Z_{PY}&=&\frac{1+\eta+\eta^{2} }{\left(1-\eta\right)^{3} }
\end{eqnarray}
Thus, given an expression for $Z$, $\chi$ can be found.  Therefore one also finds expressions for $B$ and $\mu$.  We have investigated the thermodynamics that follows from three different approximations for the compressibility factor of the steric system.  These are first,  the Carnahan-Starling expression\cite{theoryofsimpleliquids}  for $Z$
\begin{eqnarray}
Z_{CS}&=&\frac{1+\eta+\eta^{2} -\eta^{3} }{\left(1-\eta\right)^{3} },
\end{eqnarray}
second, a high density expression for $Z$, due to Mulero, Gal\'{a}n, and Cuadros\cite{Cuadros}
\begin{eqnarray}
Z_{MGC}&=&\frac{1}{\xi}\left[s+1.96192 s^{2} +0.55927 s^{3} -1.10721 s^{4} +0.55626 s^{5} \right. \\
&&\left.  -0.11923 s^{6} + 0.00954 s^{7} \right]\\
s&=&\frac{\xi}{1-\xi}\\
\xi&=&\eta/\sqrt{2},
\end{eqnarray}
and finally the Salsburg-Wood\cite{SalsburgWood} expression
\begin{eqnarray}
Z_{SW}&=&\frac{3}{\rho_{RCP}/\rho-1} + 1
\end{eqnarray}
with  the random closed packed density, $\rho_{RCP}^{*} $, = 1.21658\cite{Bernal, Scott}. While the Carnahan-Starling equation of state gives accurate values for the free energy at moderately high densities, this is not true for the other two equations of state used in this work.  Therefore, we added a small correction to the free energy in order to bring our calculated values of the free energy in agreement with calculations tabulated by Carnahan and Starling\cite{CarnahanStarling}.  Denoting the correction $\delta f$, we used $\delta f = 0$ for $Z_{CS}$, $\delta f = +.073\ N\ k_{B}$ for $Z_{MGC}$, and $\delta f = +0.413\ N\ k_{B}$ for $Z_{SW}$.
The expressions for $Z$ also give the free energy for the liquid via
\begin{eqnarray}
\frac{\beta A}{N}\equiv f_{liq} &=& \ln\rho\Lambda^{3} - 1+\int_{0}^{\eta}\left(Z-1\right)\frac{d\eta'}{\eta'}+\frac{\rho}{2}\int d\mathbf{r}g(r)\beta w(r) + \delta f
\end{eqnarray}

\subsection*{Thermodynamics of a Single Glassy Configuration}
We use self-consistent phonon theory to describe the free energy of an individual glassy configuration\cite{Fixman1969,Wolynes1984}.  This theory relies on the fact that the time averaged density can be written as a sum of gaussians representing vibrations of atoms about fiducial sites:
\begin{eqnarray}
\rho(r)&=&\sum_{i} \left(\frac{\alpha_{i} }{\pi}\right)^{3/2} \exp(-\alpha_{i} (r-R_{i} )^{2} )
\end{eqnarray}
In the present work, the force constants $\alpha_{i} $ are all taken to be equal, but this approximation need not be made, if snapshots of liquid configurations are available.  In the independent oscillator version of the theory, the effective interaction between two atoms is given by
\begin{eqnarray}
\exp(-\beta V^{eff}(r-R'))&=&\left(\frac{\alpha}{\pi}\right)^{3/2}\int d\mathbf{r'}\exp\left(-\frac{\beta}{2}V(r-r')\right)\exp\left(-\alpha (r'-R')^{2} \right)
\end{eqnarray}
Making a Taylor series expansion for the effective interaction, $\alpha$ is found self-consistently using the relation
\begin{eqnarray}
\alpha &=&\frac{\rho}{6}\int_{\Omega} d\mathbf{R}g(R)\nabla \beta V^{eff}(R)
\end{eqnarray}
where $\Omega$ is the volume of a cell containing an atom.  $\alpha = 0$ is always a solution to the self-consistent equation corresponding to a uniform liquid. Above a critical density $\rho_{A} $, the self-consistent equation gives $\alpha \ne 0$ solutions that correspond to a glassy state.  For periodic crystalline solids, Ree\cite{Ree1970} compared the uncoupled oscillator approximation with and without the Taylor series expansion for an FCC solid.  Without making the Taylor series expansion, Ree found a single solution with a value of $\alpha$ nearly zero (a "liquid-like" value)  at low densities and  the appearance of a second solution with a value of $\alpha$ characteristic of a localized state (a "solid-like" value) above a density of $\approx 1.025$.  At even higher values of the density the "liquid-like" value disappeared.  Ree interpreted the appearance of the larger value of $\alpha$ as the onset of a stable solid phase.  With the cell constraint, when  the Taylor series expansion is made only a single solution is found that continuously increases with density from a liquid-like value to a large solid-like value.  The transition from liquid-like to solid-like value occurred at approximately the same density as the appearance of the solid-like value when solving the complete  equations.  We note that  the density 1.025 is close to the values often quoted from simulation for the glass transition density (0.99-1.01\cite{Woodcock1976}, 1.05\cite{Wainwright1970}). In the present work, we found the Taylor series approximation convenient and therefore took the value for $\rho^{*}_{A} \equiv \rho_{A} \sigma^{3} $ ($=1.025 $) without the Taylor series expansion as the value we used for $\rho^{*}_{A} $. 

The free energy from the self-consistent phonon approximation is given by
\begin{eqnarray}
 f_{HS\ glass}\equiv\frac{\beta A_{HS\ glass}}{N}&=&\rho\int d\mathbf{R} g(R) \beta V^{eff}(R) + \frac{3}{2}\ln\frac{\alpha\Lambda^{2} }{\pi} - 3\ln\left[\Phi(\sqrt{\alpha} D)\right])\\
D&=&\frac{\rho^{1/3}}{2}
\end{eqnarray}
where $\Omega = D^{3} $ and we have assumed a  cubic cell for convenience.  Barker demonstrated\cite{Barker1975} that the self-consistent phonon approach produces an error of  -0.723 $N\ k_{B} $ (-0.224 $N\ k_{B} $) in the entropy of the hard sphere crystal near FCC close packing using the uncoupled (coupled) oscillator approximation.  For glasses, we expect the error in the entropy to be less than the one found for the FCC lattice and therefore we use  the somewhat smaller  correction to the coupled oscillator value for the glass. This correction has  a magnitude of 0.224 $N\ k_{B} $.  Thus, the expression we use for the free energy of an individual  Lennard-Jones glassy configuration is
\begin{eqnarray}
 f_{glass}&=&\rho\int d\mathbf{R} g(R) \beta V^{eff}(R) + \frac{3}{2}\ln\frac{\alpha\Lambda^{2} }{\pi} - 3\ln\left[\Phi(\sqrt{\alpha} D)\right])-0.224\\
&&+\frac{\rho}{2}\int d\mathbf{r}g(r)\beta w(r)
\end{eqnarray}
With this correction, the theory yields values of $\rho^{*}_{K} $ ranging from 1.16 to 1.24, depending on the equation of state used to determine $g(r)$.  For comparison, using a liquid structure based density functional theory, Singh, Stoessel and Wolynes\cite{Wolynes1985} found $\rho^{*}_{K} = 1.14$.  

\subsection*{Pressure, Configurational Entropy, and Heat Capacities}
Random first order transition theory identifies the configurational entropy with  the difference between the free energies of the glass and the liquid.  Thus, the configurational entropy is
\begin{eqnarray}
TS_{c} & = &A_{glass}-A_{liquid}\\
S_{c} / N\ k_{B} & = & f_{glass}-f_{liquid}\equiv \Delta f
\end{eqnarray}

With the present liquid state approximation, the best fits of $\Delta f$ versus temperature were obtained using the function
\begin{eqnarray}
\Delta f &=& a_0 + a_{1} \ln T +\frac{a_{2} }{T} + \frac{a_{3} }{T^{2} } + \frac{a_{4} }{T^{3} }
\end{eqnarray}
$\Delta C_{p} $ and $\Delta C_{V} $ were then be found by differentiation:
\begin{eqnarray}
\Delta C_{V} /N\ k_{B} &=&T\left(\frac{\partial \Delta f}{\partial T}\right)_{V} \label{CVder}\\
\Delta C_{p} /N\ k_{B} &=&T\left(\frac{\partial \Delta f}{\partial T}\right)_{p} 
\end{eqnarray}
The fitting function has the flexibility to describe heat capacities that depend on temperature as either $1/T$ or $1/T^{2} $; both functional dependencies have been suggested\cite{Angell1972,Angell2005}  in empirical and model studies. 
Within the liquid state theoretic incarnation of random first order transition theory, we see that no physical significance can be ascribed to these fits such as "Gaussianity of the landscape" or inferences of "excitation" structure as in some picturesque models. The temperature dependence simply reflects the softening of the steric potentials with increased kinetic energy of the molecules.

The pressure can be evaluated by numerically differentiating the liquid free energy:
\begin{eqnarray}
p&=&\rho^{2} \left(\frac{\partial f}{\partial \rho}\right)_{T,N}k_{B} T
\end{eqnarray}

\section{Results from Simulations}
To orient the reader  to the present study, we first describe computer simulations of simple glasses which have previously been carried out. Computer simulations are often used to discuss the structure and dynamics of glassy materials.  Yet,  despite their widespread use, carrying out meaningful simulations of these systems is not easy.  Direct studies are hindered by several factors.  One difficulty is the need to avoid crystallization.  The other challenge is  to reach the very long time scales consistent with laboratory experiments  (microsecond to seconds).  So far this has never been done. For monatomic glasses, even on a short time scale avoiding crystallization is extremely difficult.  This is one of the reasons binary Lennard-Jones mixtures have become popular choices for simulation. When crystallization is avoided it is still important to recognize that $T_{g}$ depends on cooling rate, albeit logarithmically.
J\'{o}nsson and Andersen\cite{Andersen1988} found $T^{*}_{g} \equiv k_{B} T_{g} /\epsilon = .38$ in a constant pressure simulation of an 80/20 mixture (with parameters chosen similar to those used by Stillinger and Weber to study amorphous $Ni_{80}P_{20}$\cite{Stillinger1985}) with a cooling rate of $\Delta T^{*} /\Delta t^{*} = 0.001$.  Kob and Andersen\cite{Andersen1995} used smaller cooling rates (as slow as $1.5 \times 10^{-7}$),  fit the diffusion constant to the predictions of mode coupling theory, and determined $T^{*}_{MCT} = 0.435$. Sastry, Debenedetti, and Stillinger\cite{Stillinger1998} used cooling rates ranging from $10^{-4}$ to $10^{-7}$ 
 and found the onset of activated dynamics at $T^{*} < 0.45$ at a density $\rho^{*} = 1.2$. Sastry fit the temperature dependence of the diffusion constant calculated in simulations.
 The fits give  $T^{*}_{K} = 0.30$ at $\rho^{*} = 1.2$, similar to values found by Sciorino, Kob, and Tartaglia from an analysis of inherent structures\cite{Tartaglia1999} and by Coluzzi, Parisi, and Verrocchio\cite{Verrocchio2000}, who used a replica-based approach.  

Many simulations study monatomic Lennard-Jones systems but modify the  Lennard-Jones potential by adding  a small many-body potential energy term that discourages  crystallization\cite{Sastry1997}.  The modified potential  has triple point and critical point values $T^{*}_{t.p.}=0.687, p^{*}_{t.p.}\equiv p_{t.p.}\sigma^{3} / \epsilon =0.0030619, \rho^{*}_{t.p.}=0.67, T^{*}_{c}=1.16, p^{*}_{c} = 0.109, $ and $ \rho^{*}_{c} = 0.247$\cite{Sastry1997}, which can be compared to the values for the full LJ potential, $T^{*}_{t.p.} = 0.67, \rho^{*}_{f, t.p.} = 0.86, T^{*}_{c} = 1.25-1.58, p^{*}_{c} = .109-.303, $ and $\rho^{*}_{c} = 0.26-0.40$\cite{theoryofsimpleliquids}.  Fits of the simulated diffusion constant to the results of mode coupling phenomenology,  $(T^{*}-T^{*}_{MCT})^{\gamma}$ give $T^{*}_{MCT} = 0.475 $ at $\rho^{*} = 1.0 $\cite{Sciortino2002}.  Calculations using the modified Lennard-Jones potential and a cooling rate of $4.2 \times 10^{-4}$  found  $0.3 < T^{*}_{g} < 1.13$ for densities $0.85 < \rho^{*} < 1.25$\cite{Ruocco2000}.  

Clearly the cooling rates for the present day simulations are still  much faster than those used in experiments.  According to Angell\cite{Angell2002}, laboratory cooling rates of $\approx$ 0.17 K/s give the experimental glass transition at the temperature at which the structural relaxation time equals 100 seconds. With some cleverness cooling rates as fast as $10^{6} $ K/s have been reached for melt-spun glasses\cite{Angell2002}.  In dimensionless simulation units cooling rates as small as  $10^{-7} = dT^{*} / dt^{*}  $ and as large as $10^{-3} = dT^{*} / dt^{*}$ have been used in Lennard-Jones simulations, where $t^{*} = t\left(\frac{48 \epsilon }{m \sigma^{2} }\right)^{1/2}$. Rescaling the  units for argon, using $\epsilon/k_{B} = 120$ K and $\sigma = 3.4$ A, the range of simulated cooling rates is then $10^{7} < \Delta T^{*} /\Delta t^{*} < 10^{11}$ K/s.  For the monatomic Lennard-Jones simulation\cite{Ruocco2000}, therefore, the cooling rate is roughly $10^{4} $ times faster than the fastest experimental rate.  As a result, values of $T^{*}_{g} $ obtained from  simulation must be viewed as upper limits to the laboratory glass transition.  Velikov, Borick, and Angell\cite{Angell2002} estimate that simulations overestimate $T^{*}_{g} $ by a factor of 1.2-1.6.

As the temperature of liquid is increased at fixed density, the spinodal temperature $T_{spinodal}^{*} $ represents the limit of stability of the liquid with respect to formation of a gas.  Similarly, $T^{*}_{K} $ represents the lowest temperature at which a mobile liquid can be considered stable with respect to an "ideal" glass\cite{DiMarzio1958,Gibbs1965,Speedy2004}.  Sastry, Debenedetti, and Stillinger\cite{Sastry1997} determined the distributions of "voids", inhomogeneities in the liquid, in simulations of the modified Lennard-Jones glass.  Voids in the frozen glass appeared only below $\rho^{*} = 0.89$, which appeared to be the $T=0$ extrapolation of the liquid spinodal.  Sastry examined this phenomena more closely\cite{Sastry2000} and suggested that  the spinodal and Kauzmann temperatures intersect at $T^{*} = 0.16 $ at $\rho^{*} = 1.08$ for a binary Lennard-Jones mixture.  

\section{Numerical Results and Comparison to Simulations}

\begin{figure}[h]
\begin{center}
\includegraphics[width=6in]{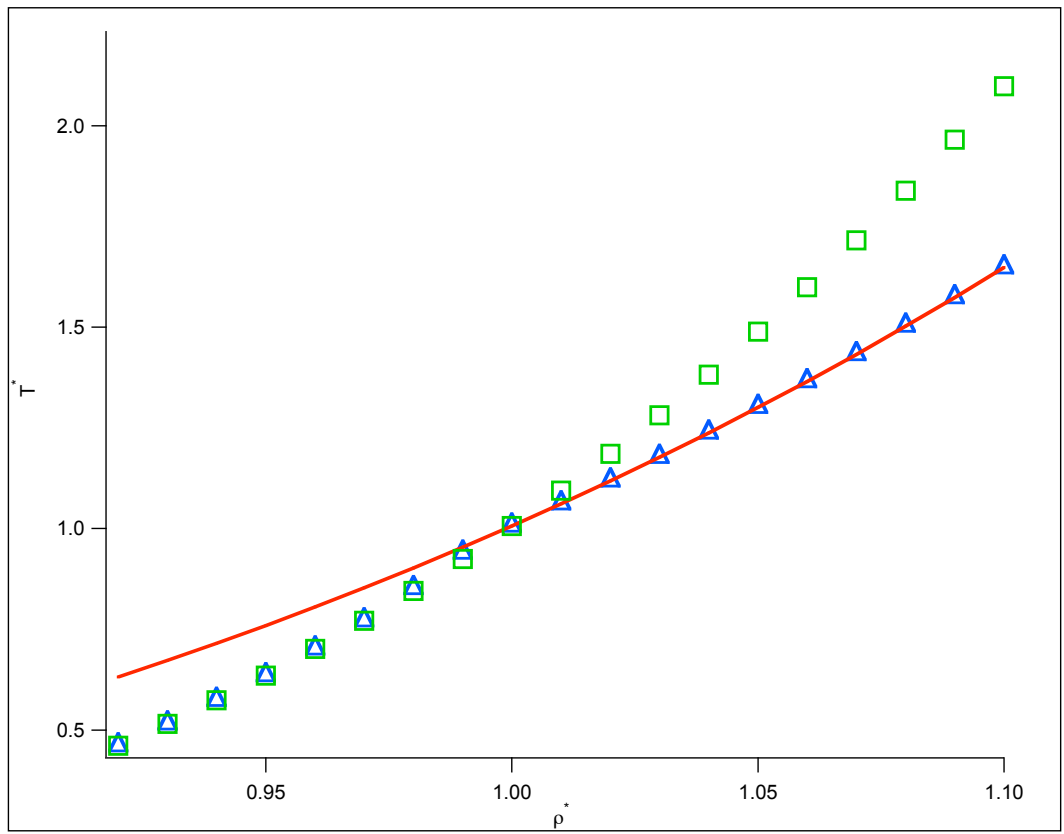}
\caption{$T^{*}_{A} $  as a function of density using the MGC equation of state\cite{Cuadros} and the WCA (squares), KRR (triangles), and KRR with constant $\lambda$ (solid line) separations. \label{TAWCAvsKRR}}
\end{center}
\end{figure}

$T^{*}_{A} $, $T^{*}_{c}$, $T^{*}_{g} $, and $T^{*}_{K} $ were calculated for several densities $\rho^{*}$.  We determined the mean field stability limit $T^{*}_{A} $ as the temperature at which $\rho_{A} d^{3} = 1.025$, where $d$ is the effective hard sphere diameter determined from Eqn.~\ref{blip}.  The crossover temperature, accounting for fractal excitations, $T^{*}_{c}$ was determined by the condition that $\Delta f = 1.28$. $T^{*}_{K} $ was determined as the temperature at which the configurational entropy vanishes, i.e., $\Delta f = 0$. Following Lubchenko and Wolynes\cite{Wolynes2003}, the laboratory glass transition temperature  $T^{*}_{g} $ was  calculated using Eqn.~\ref{timescale} for the  one hour time scale, therefore $T^{*}_{g}$ corresponds to $\Delta f = 0.82$.  Figure~\ref{TAWCAvsKRR} displays $T^{*}_{A} $ as a function of  density for the  Mulero-Gal\'{a}n-Cuadros  equation of state described above and for the WCA and KRR separations, as well as for a modification of the KRR separation in which $\lambda = a(\rho)$ at all densities (the reasons for this latter separation are discussed below).  The results obtained for $T^{*}_{A}$, $T^{*}_{c}$, $T^{*}_{g} $, and $T^{*}_{K} $ using the other equations of state were qualitatively similar to each other. As expected, at the lower densities ($\rho^{*} < \approx 1.01$), the KRR and WCA separations give essentially the same results.  At higher densities, ($\rho^{*} > \approx 1.01$), however the results from the  different potential separations are quantitatively different. Because of its discontinuous definition, the KRR separation as it stands would give  a rather unnatural looking change in slope in the region around ($\rho^{*} \approx 1.01$).  We therefore also explored the predictions of the KRR theory when one chooses $\lambda = a(\rho)$ at all densities. The corresponding results for the KRR separation are shown as solid lines in the figure.  All three equations of state give $T_{A} ^{*}  \approx 1.0 $ at $\rho^{*} = 1.0$. 

It is important to recognize that the stability limit computed from self-consistent phonon theory corresponds to a strictly mean field mode coupling limit. Our  value of $T^{*}_{A} $  corresponds well with that found from detailed mode coupling calculations\cite{Sciortino2002}. As Reichmann and coworkers have pointed out\cite{Reichman2004}, structure based mode coupling calculations give a larger value of mode coupling temperature than do fits to simulation data using phenomenological mode coupling expressions. The latter phenomenological fits\cite{Sciortino2002} yield a lower temperature $T^{*}_{MCT} = 0.475$. This discrepancy between structure based mean field calculations and phenomenology has been explained by Stevenson, Schmalian and Wolynes\cite{SchmalianWolynes2006}. They have shown that fractally shaped excitations allow downhill escape from a mean field minima at temperature below the mean field $T^{*}_{A}$.  Thus, they argue that when fractal or string excitations are allowed in random first order transition dynamics, the true stability limit of an aperiodic structure is suppressed below its mean field estimate. The crossover to activated dynamics in their analysis occurs at a configurational entropy value of $S_{c} / N\ k_{B}  = 1.28$, according to the simpler estimates based on percolation. We designate this crossover temperature as $T^{*}_{c} $   and compute it as well as the mean field $T^{*}_{A} $. Its value at $p = 0.0$ is $T^{*}_{c} = 0.35$. This crossover temperature agrees better with the fits of mode coupling phenomenology to simulation. 

\begin{figure}[htbp]
\begin{center}
\includegraphics[width=6in]{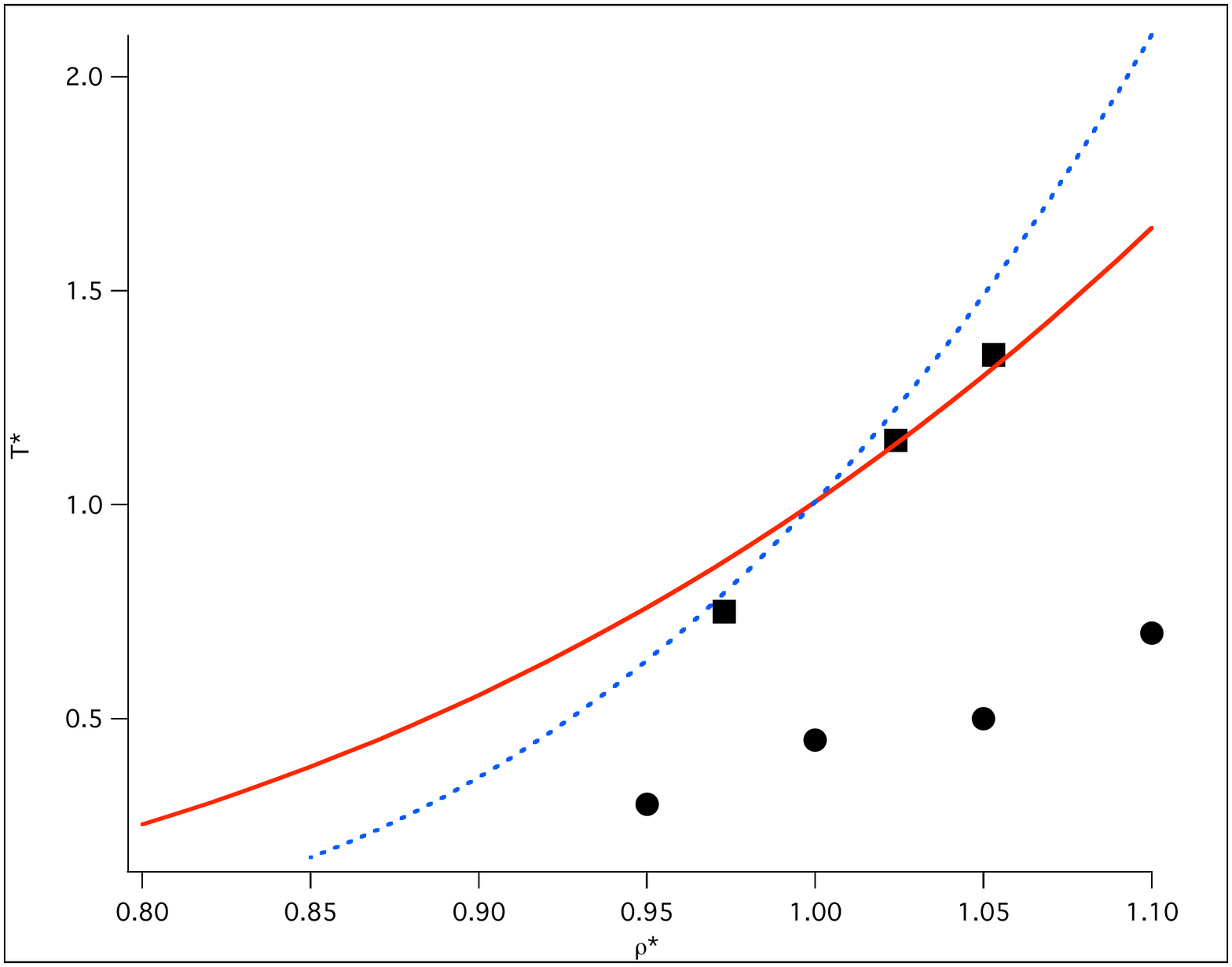}
\caption{$T^{*}_{A} $ as a function of density using  $Z_{MGC}$ for the modified ($\lambda = a(\rho)$) KRR (solid line) and WCA (dashed line) separations.  $T^{*}_{m} $, the melting temperture\cite{Verlet1969} (filled squares) and $T^{*}_{g,sim}$, the glass transition temperature at the simulation time scale\cite{Ruocco2000} (filled circles) are also shown.\label{MGCTA}}
\end{center}
\end{figure}

\begin{figure}[htbp]
\begin{center}
\includegraphics[width=6in]{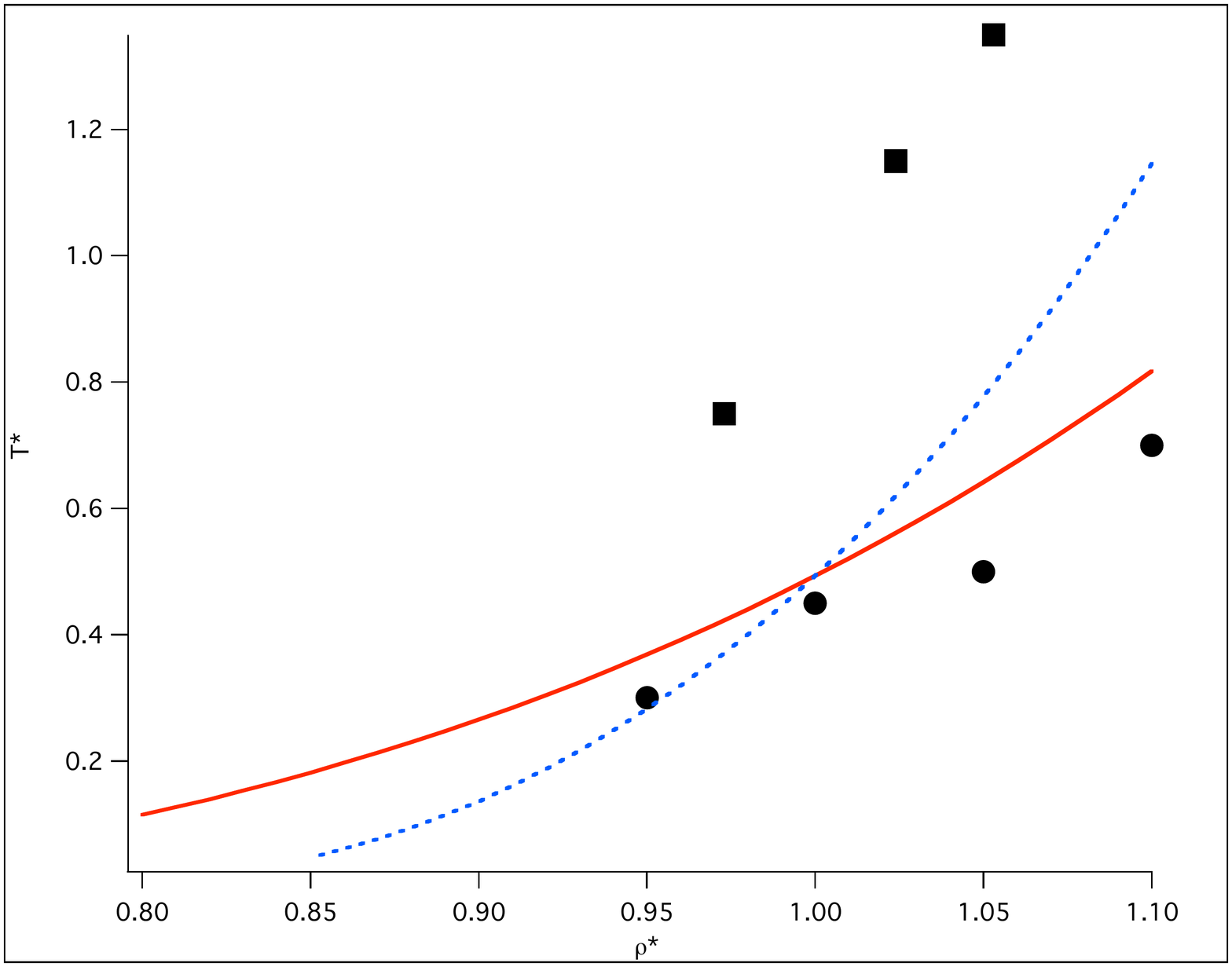}
\caption{The crossover temperature $T^{*}_{c} $ as a function of density using  $Z_{MGC}$ for the modified ($\lambda = a(\rho)$) KRR (solid line) and WCA (dashed line) separations.  $T^{*}_{m} $, the melting temperture\cite{Verlet1969} (filled squares) and $T^{*}_{g,sim}$, the glass transition temperature at the simulation time scale\cite{Ruocco2000} (filled circles) are also shown.\label{MGCTC}}
\end{center}
\end{figure}

\begin{figure}[htbp]
\begin{center}
\includegraphics[width=6in]{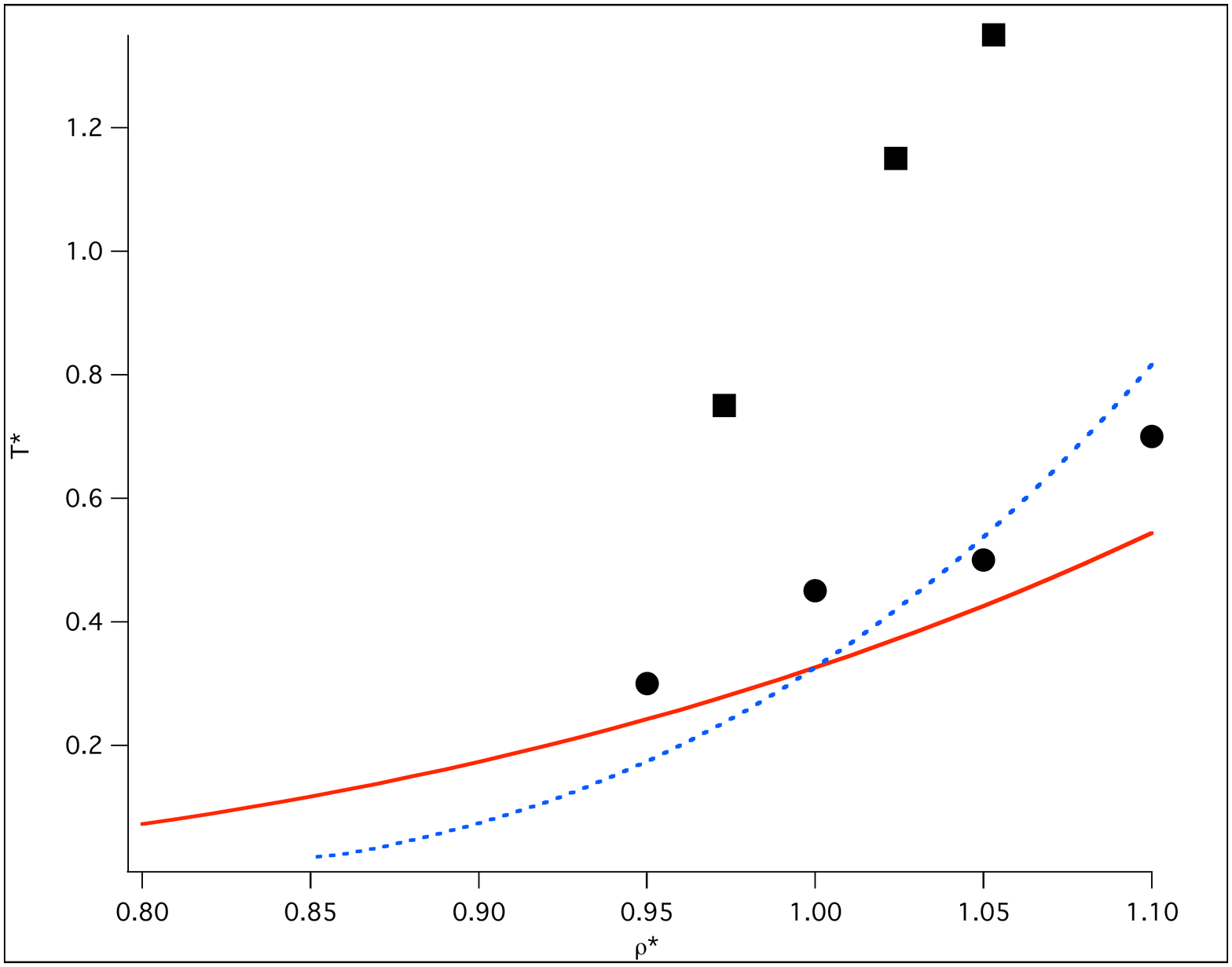}
\caption{$T^{*}_{g} $ as a function of density using  $Z_{MGC}$ for the modified ($\lambda = a(\rho)$) KRR (solid line) and WCA (dashed line) separations.  $T^{*}_{m} $, the melting temperture\cite{Verlet1969} (filled squares) and $T^{*}_{g,sim}$, the glass transition temperature at the simulation time scale\cite{Ruocco2000} (filled circles) are also shown.\label{MGCTG}}
\end{center}
\end{figure}

\begin{figure}[htbp]
\begin{center}
\includegraphics[width=6in]{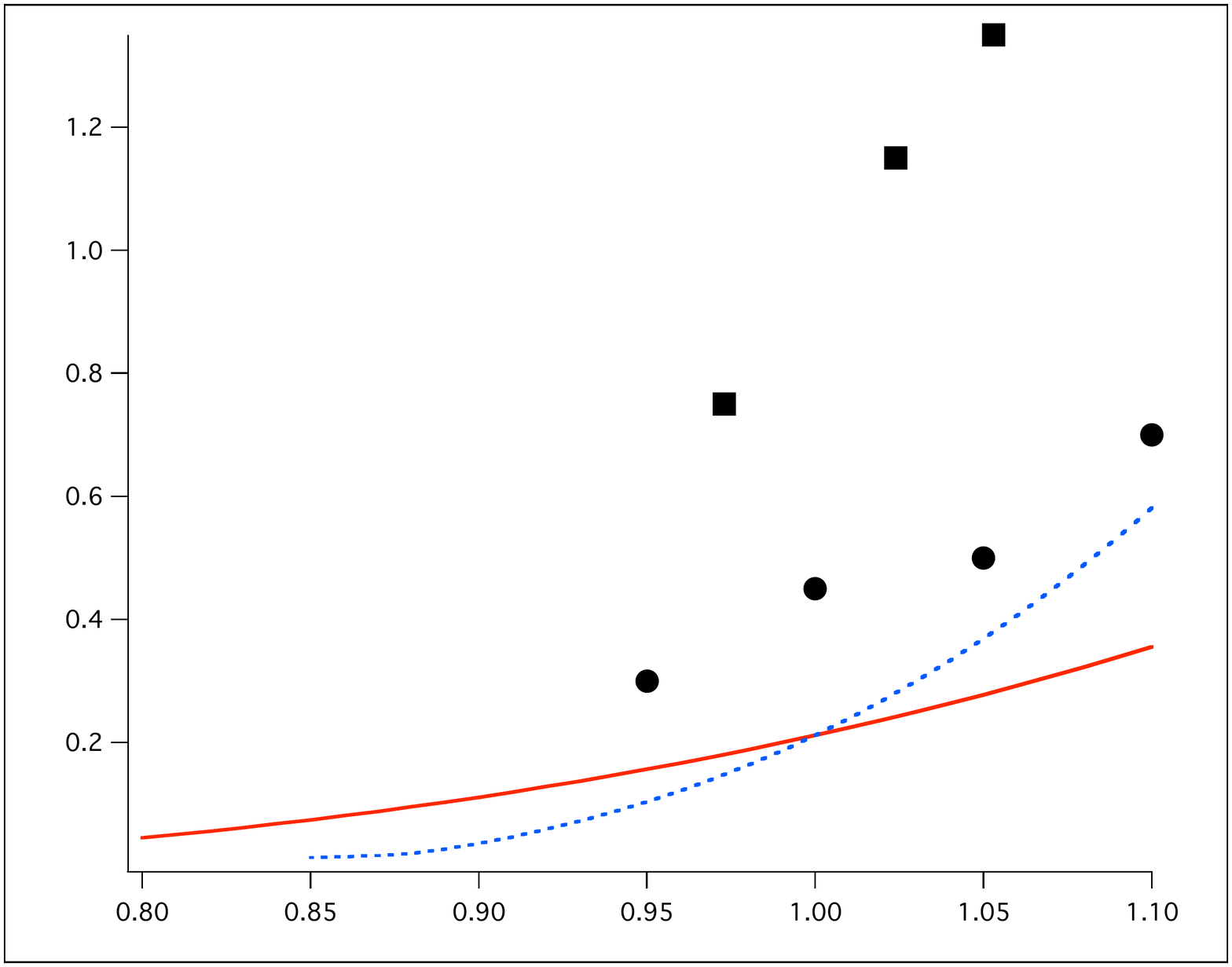}
\caption{$T^{*}_{K} $ as a function of density using  $Z_{MGC}$ for the modified ($\lambda = a(\rho)$) KRR and WCA separations.  $T^{*}_{m} $, the melting temperture\cite{Verlet1969} (filled squares) and $T^{*}_{g,sim}$, the glass transition temperature at the simulation time scale\cite{Ruocco2000} (filled circles) are also shown.\label{MGCTK}}
\end{center}
\end{figure}

Figures~\ref{MGCTA}-\ref{MGCTK} display the values of $T^{*}_{A} $, $T^{*}_{c} $, $T^{*}_{g} $, and $T^{*}_{K} $, respectively, predicted by our theory using the Mulero-Gal\'{a}n-Cuadros equation of state and both the modified KRR and WCA perturbation approaches. The Salzburg-Wood equation of state gives nearly identical values for these quantities, while the Carnahan-Starling equation of state gives significantly lower values for all the temperatures except for $T^{*}_{A} $.  Also shown are values from simulation\cite{Verlet1969} for the melting temperature, $T^{*}_{m} $ and the glass transition temperature\cite{Ruocco2000} on the simulation time scale, $T^{*}_{g, sim}$. 
 Examination of our results for $T^{*}_{c} $ and $T^{*}_{g} $ show that the modified KRR separation is in better agreement with simulation than is the WCA separation.  Therefore, in the remainder of this paper we have used the modified KRR separation.  
 
The values of $T_{A}^{*}$ are close to the melting temperature at high densities.  $T^{*}_{g, sim}$ lies between the values of $T^{*}_{c} $ and $T^{*}_{g} $, as expected.  We find $T^{*}_{g, sim} / T^{*}_{g}  \approx 1.3$ close to the estimate of  Angell\cite{Angell2002}.  The present theory gives $T^{*}_{c} / T^{*}_{g} = 1.5$.
Stevenson and Wolynes examined\cite{Stevenson2005}  experimental values for $T_{g} $ and confirmed  the well-known rule of thumb that $T_{g} / T_{m} \approx 2/3$.  The present calculations yield lower ratios, $T^{*}_{g} / T^{*}_{m} \approx 0.4 $ for $Z_{SW}$ and $Z_{MGC}$ and $\approx 0.35 $ for $Z_{CS}$.

\begin{figure}
\begin{center}
\includegraphics[width=6in]{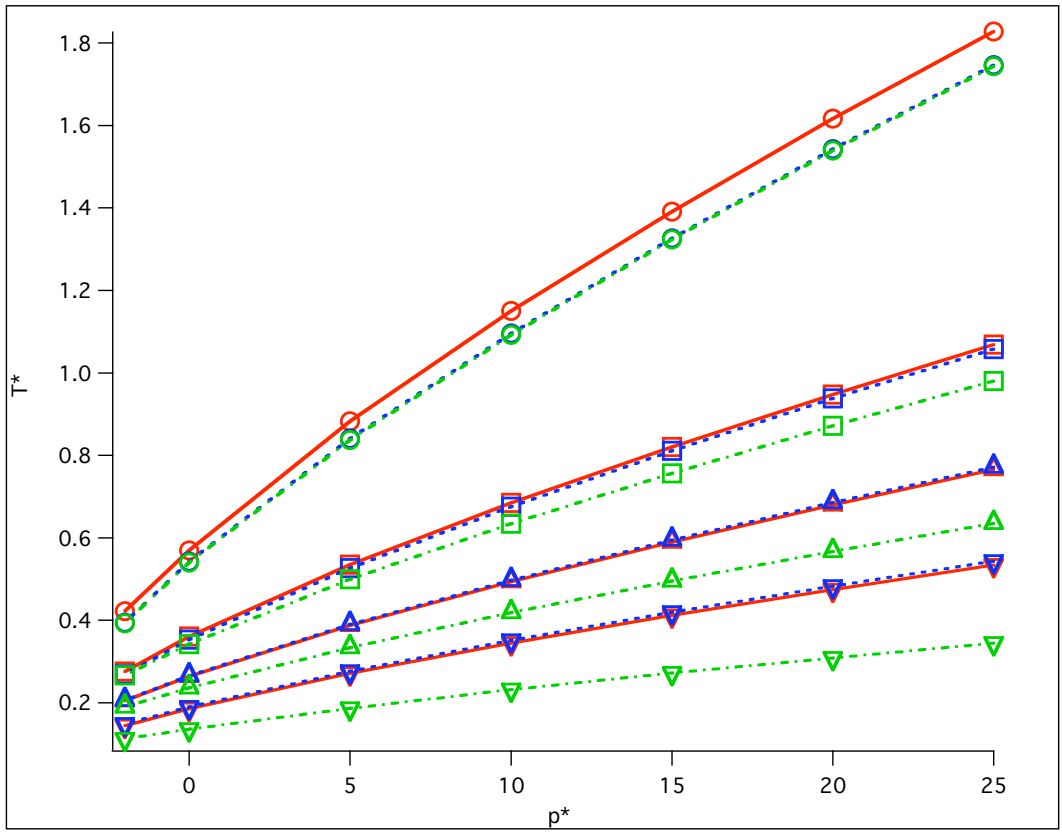}
\caption{$T^{*}_{A} $ (circles), $T^{*}_{c}$ (squares), $T^{*}_{g}  $ (triangles)  , and $T^{*}_{K}  $ (inverted triangles) versus $p^{*}$ predicted using $Z_{SW}$ (solid lines), $Z_{MGC}$ (dashed lines), and $Z_{CS}$ (dash-dot lines).  The results using $Z_{SW}$ and $Z_{MGC}$ are almost identical.\label{ratiospconstant}}
\end{center}
\end{figure}

Figure~\ref{ratiospconstant} displays $T^{*}_{A}  $, $T^{*}_{c} $, $T^{*}_{g} $, and $T^{*}_{K}  $ at various values of  the  pressure.  Angell\cite{Angell1997} has tabulated values of $T_{K} / T_{g} $ from experiment and has found the ratio to typically be $\approx 0.78$ for fragile substances at atmospheric pressure.   The values from the present theory at $p^{*} = 0$ are $T^{*} _{K} / T^{*} _{g} = 0.70, 0.72, 0.58$, for $Z_{SW}$, $Z_{MGC}$, and $Z_{CS}$, respectively.  While the results from all equations of state agree at the higher temperatures displayed, the results based on the Carnahan-Starling equation of state are substantially different than the other equations of state at lower temperatures.  The good agreement using equations of state appropriate to high density with previous simulations demonstrates the ability of the present theory to describe the different temperature regimes in the Lennard-Jones glass.

\begin{figure}
\begin{center}
\includegraphics[width=6in]{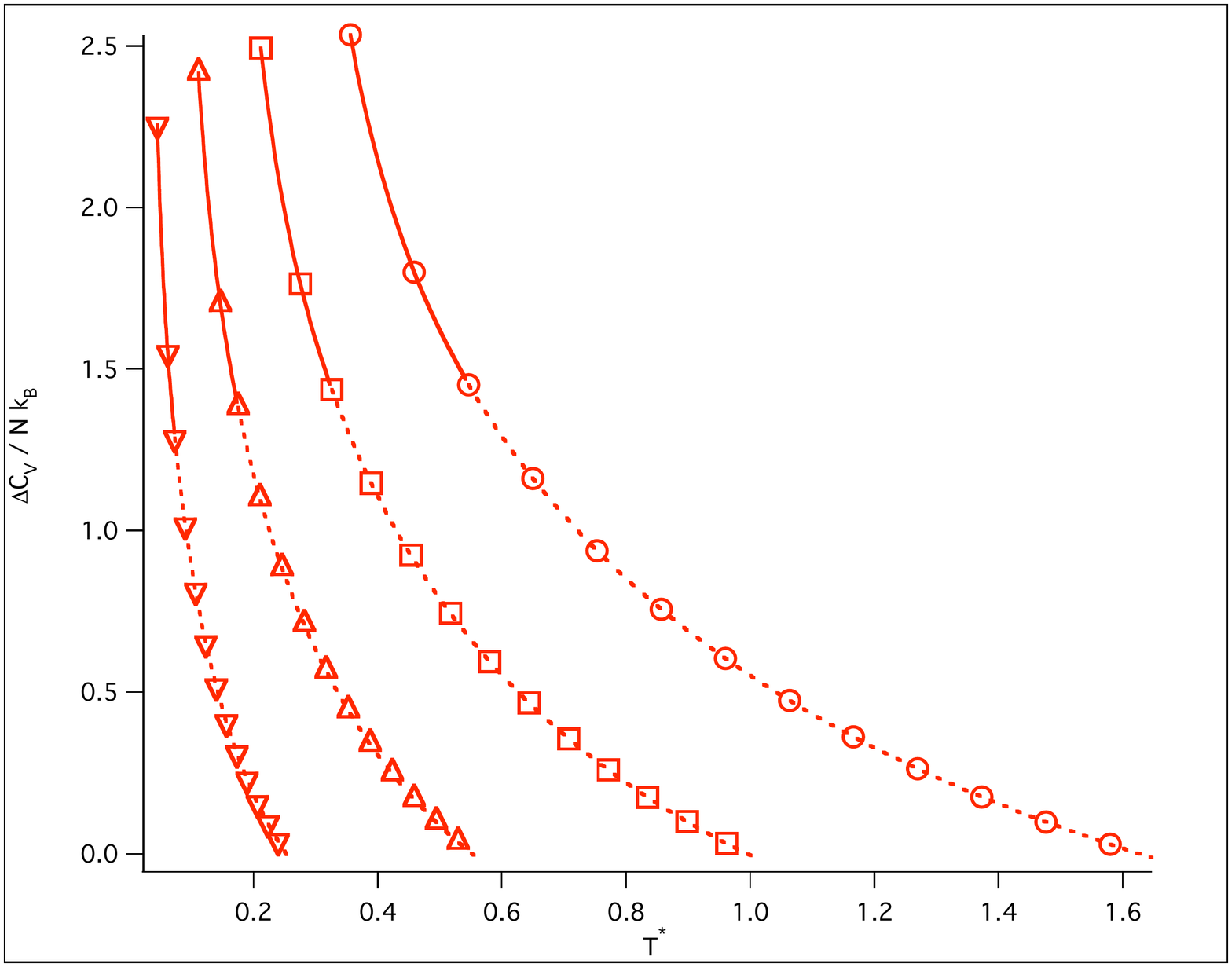}
\caption{$\Delta C_{V} /N\ k_{B}$ versus $T^{*} $ using $Z_{MGC}$ at densities $\rho^{*} = 1.10$ (circles), $\rho^{*} = 1.00$ (squares), $\rho^{*} = 0.90$ (triangles), and $\rho^{*} = 0.80$ (inverted triangles).  The curves use solid (dashed) lines for temperatures below (above) $T^{*}_{g}$.\label{CV1}}
\end{center}
\end{figure}

\begin{figure}
\begin{center}
\includegraphics[width=6in]{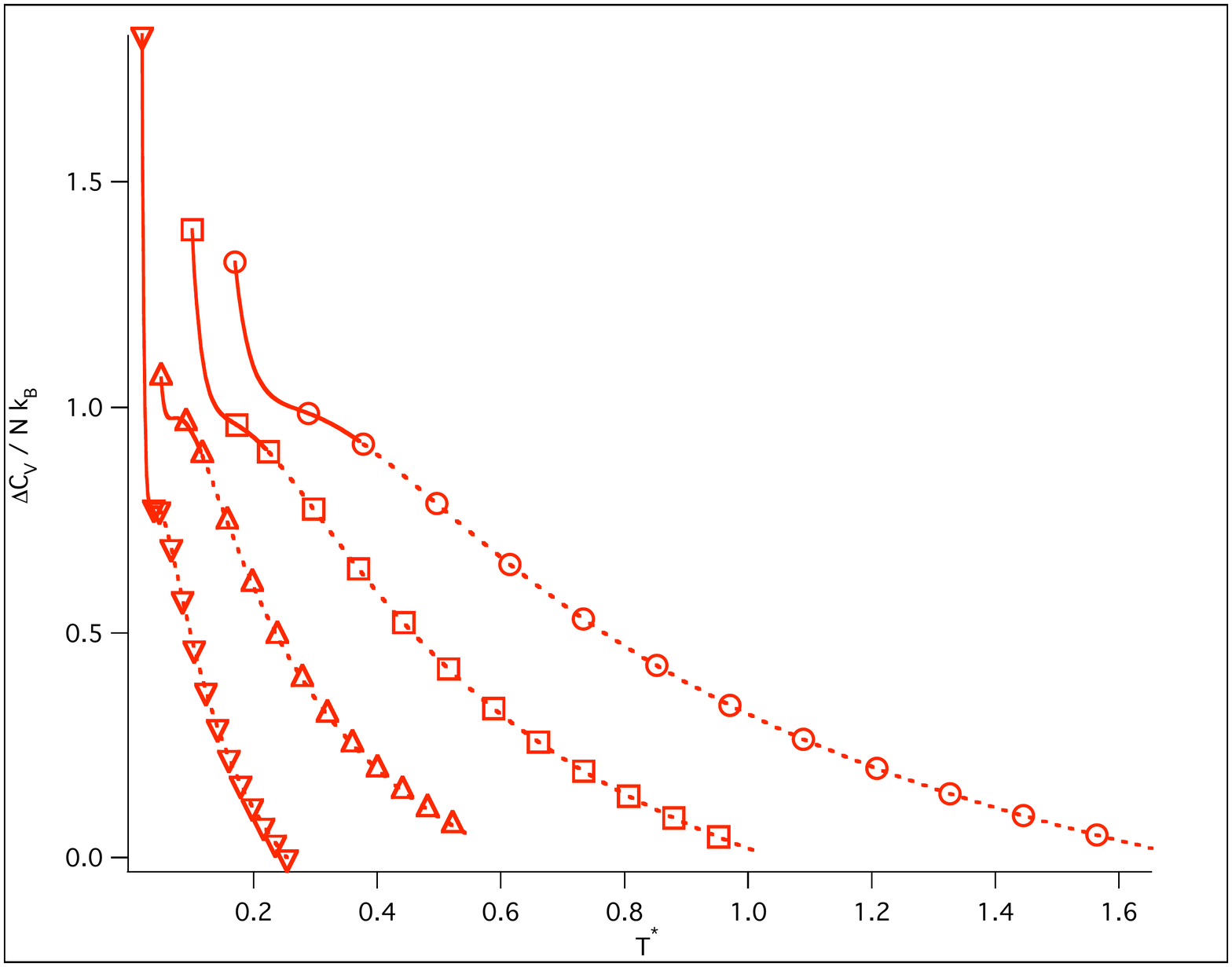}
\caption{$\Delta C_{V} /N\ k_{B}$ versus $T^{*} $ using $Z_{CS}$ at densities $\rho^{*} = 1.10$ (circles), $\rho^{*} = 1.00$ (squares), $\rho^{*} = 0.90$ (triangles), and $\rho^{*} = 0.80$ (inverted triangles).  The curves use solid (dashed) lines for temperatures below (above) $T^{*}_{g}$.\label{CV2}}
\end{center}
\end{figure}

\begin{figure}
\begin{center}
\includegraphics[width=6in]{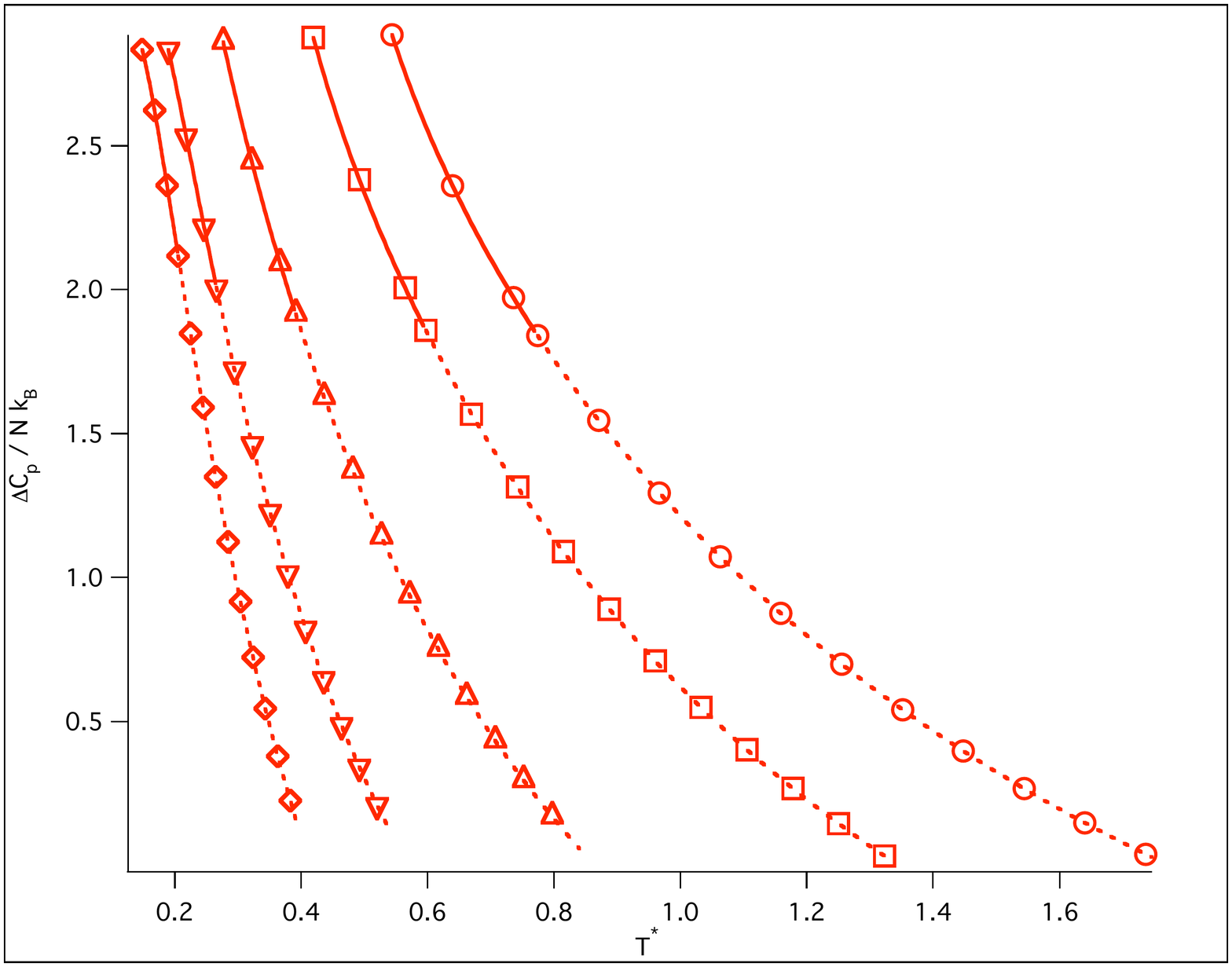}
\caption{$\Delta C_{p} $ versus $T^{*} $  using $Z_{MGC}$ at pressures $p^{*} = 25.0$ (circles), $p^{*} = 15.0$ (squares), $p^{*} = 5.0$ (triangles), $p^{*} = 0.0$ (inverted triangles), and $p^{*} = -2.0$ (diamonds). The curves use solid (dashed) lines for temperatures below (above) $T^{*}_{g}$.\label{cp1}}
\end{center}
\end{figure}

\begin{figure}
\begin{center}
\includegraphics[width=6in]{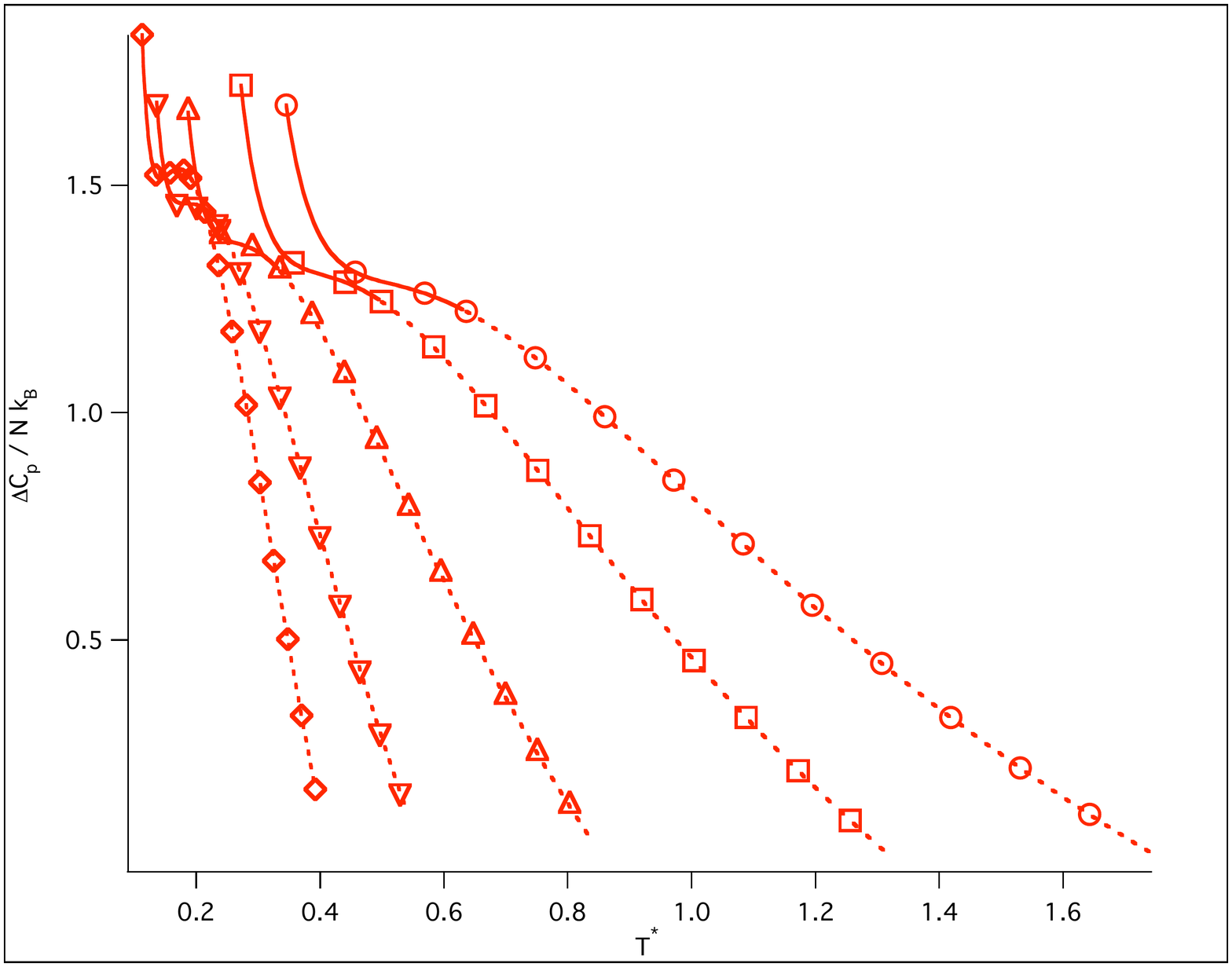}
\caption{$\Delta C_{p} $ versus $T^{*} $  using $Z_{CS}$ at pressures $p^{*} = 25.0$ (circles), $p^{*} = 15.0$ (squares), $p^{*} = 5.0$ (triangles), $p^{*} = 0.0$ (inverted triangles), and $p^{*} = -2.0$ (diamonds). The curves use solid (dashed) lines for temperatures below (above) $T^{*}_{g}$.\label{cp2}}
\end{center}
\end{figure}

The temperature dependences of the predictions for the constant volume and constant pressure heat capacity discontinuities $\Delta C_{V} $  and $\Delta C_{p} $ are displayed in Figures~\ref{CV1}-\ref{cp2}. The results found using $Z_{SW}$ (not shown in the figures) and $Z_{MGC}$ are quite similar to each other but are  distinct (particularly in their high temperature behavior) from the results found using $Z_{CS}$.  Table~\ref{cv_table} gives the values of $\Delta C_{V} /N k_{B} $ at $T^{*}_{g} $ and Table~\ref{cp_table} gives the values of $\Delta C_{p} /N k_{B} $ at $T^{*}_{g} $.  $\Delta C_{p} /N k_{B} $ has been tabulated for a variety of glasses by Stevenson and Wolynes\cite{Stevenson2005} and the average value of $\Delta C_{p} /N k_{B} $ for glasses of medium fragility  at $T^{*}_{g} $ per bead is 2.85. For the smaller molecular glasses methanol, n-propanol, butyronitrile, ethylene, and ethanol, the value per bead is somewhat smaller, averaging 2.1. The $p^{*}=0$ values for the heat capacity gaps using $Z_{MGC}$ ($\Delta C_{p} \approx 1.9$) are in agreement with those found for the smaller molecular glasses.

\begin{figure}
\begin{center}
\includegraphics[width=6in]{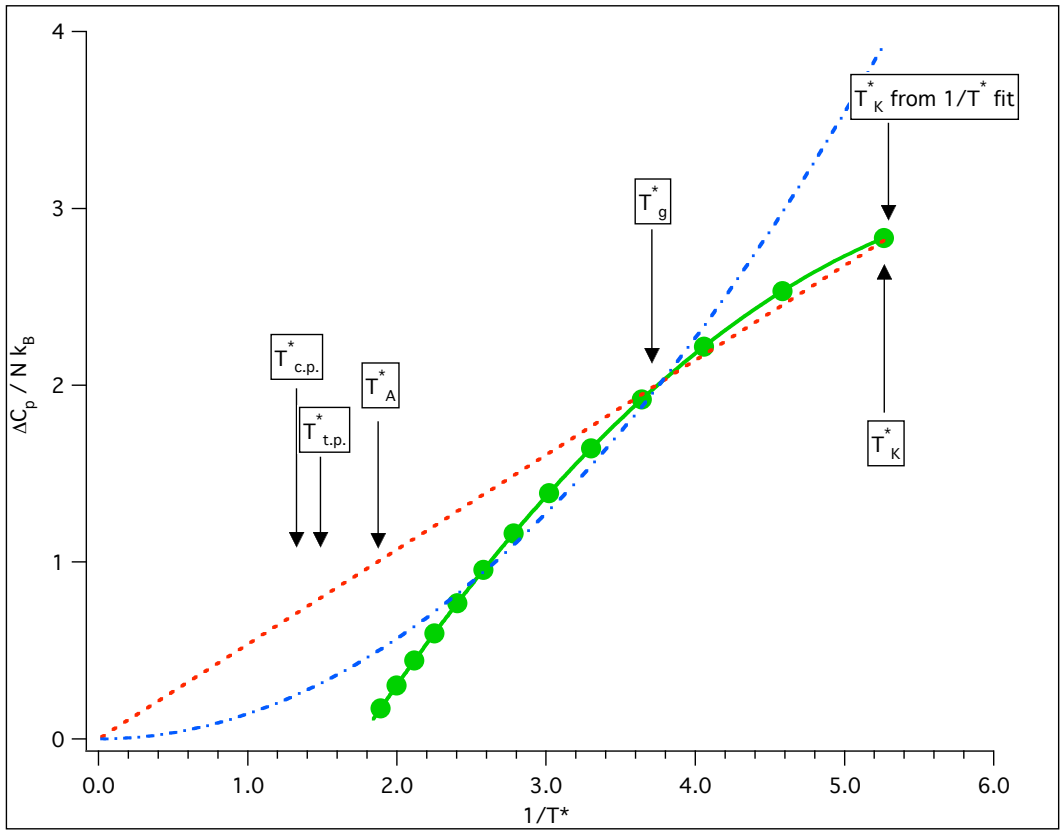}
\caption{Fits of $\Delta C_{p} $ calculated using $Z_{MGC}$ (filled circles) versus $1/T^{*} $ (dashed line) and $1/T^{*2}$ (dash-dot line).  The reduced pressure is 0.0 and the fits were performed by forcing agreement at the temperature $T^{*}_{g}$. The critical ($T^{*}_{c.p.}$) and triple ($T^{*}_{t.p.}$) temperatures from simulation\cite{Protsenko2005} are also shown, as are the values of $T^{*}_{A}$, $T^{*}_{g}$, and $T^{*}_{K}$ from our calculations.  The value of $T^{*}_{K}$ obtained by integrating the $1/T^{*}$ fit to determine the temperature at which the configuration entropy would go to zero is also indicated.\label{cpfit}}
\end{center}
\end{figure}

We investigated the conjectures that $\Delta C_{p} = k/T $ and $\Delta C_{p}= k'/T^{2}$ that are often used in fitting laboratory data.  The constants were chosen so that the fits agreed with the calculated data at the temperature $T^{*}_{g}$.  The results of the fits are shown in Figure~\ref{cpfit}.  It is clear that neither of these fitting functions accurately represents the calculated heat capacity, though the $1/T^{*2}$ fit is accurate to about 10\% above $T^{*}_{g}$ except at temperatures close to $T^{*}_{A}$.  Conversely, the $1/T^{*}$ fit is better at temperatures below the glass transition temperature and can be used determine $T^{*}_{K}$ accurately by integration.  The prediction of this procedure is shown in the figure. It is clear that the temperature dependence of the calculated heat capacities does not precisely follow the simple forms suggested by either gaussian landscapes or by elementary two state excitations.

In discussing the role of intermolecular forces and glasses the empirical correlations that have been found between the $\alpha$-relaxation time $\tau$ and $1/T V^{\gamma}$ are of keen interest. The  empirical parameter $\gamma$ varies from system to system. The fits to laboratory data give  values for $\gamma$ as small as 0.13 for sorbitol and as large as 8.5 for 1,1'-di(4-methoxy-5-methylphenyl)-cyclohexane (BMMPC).  For a variety of pressures, by varying $\gamma$ nearly universal behavior is found when $\log_{10}\tau$ is plotted versus $1/T V^{\gamma}$. For a $1/r^{12}$ soft-sphere potential energy, the configurational free energy  must be  a universal function of $\rho^{*4}  / T$ by dimensional analysis, a fact noticed by several authors\cite{Hoover1971,HallWolynes2003,Casalini2007,Casalini2007b}.  This result is consistent with many of the experimental correlations since random first order transition theory provides a link between $S_{c} $ and $\alpha$-relaxation times\cite{Wolynes2000,Wolynes2001A}.  The random first order transition theory  expression for the relaxation time $\tau =\tau_0 \exp\left(\frac{\Delta F^{\ddagger}}{k_{B} T}\right)$ and 
$\frac{\Delta F^{\ddagger}}{k_{B} T} =\frac{32}{S_{c} / N k_{B} }\label{32}$
leads to 
\begin{eqnarray}
\log_{10}\tau &=&\log_{10}\tau_0 + \frac{14 N k_{B} }{S_{c} }
\end{eqnarray}

\begin{figure}
\begin{center}
\includegraphics[width=6in]{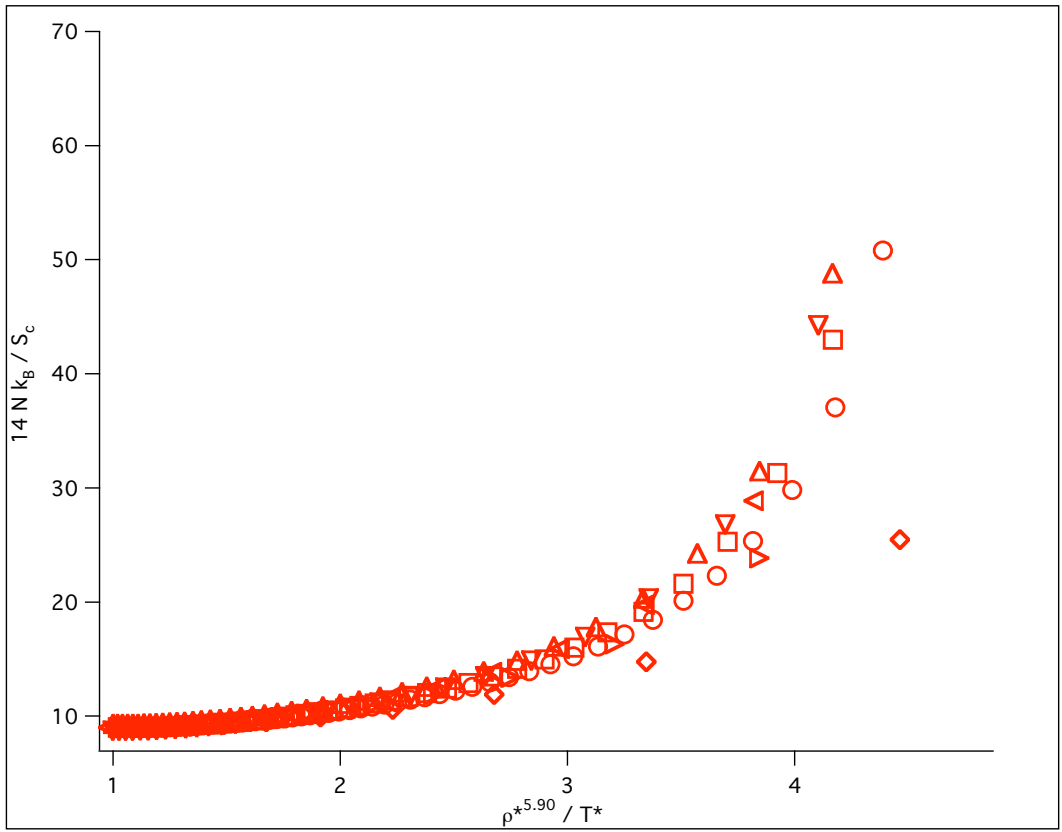}
\caption{Plots of $1/S_{c}$ versus $\rho^{*5.90} /T^{*} $ for $Z_{MGC}$.  Values of the density are 1.10 (circles), 1.05 (squares), 1.00 (triangles), 0.95 (inverted triangles), 0.90 (left-facing triangle), 0.85 (right-facing triangle), and 0.80 (diamonds).\label{casalini2}}
\end{center}
\end{figure}

\begin{figure}
\begin{center}
\includegraphics[width=6in]{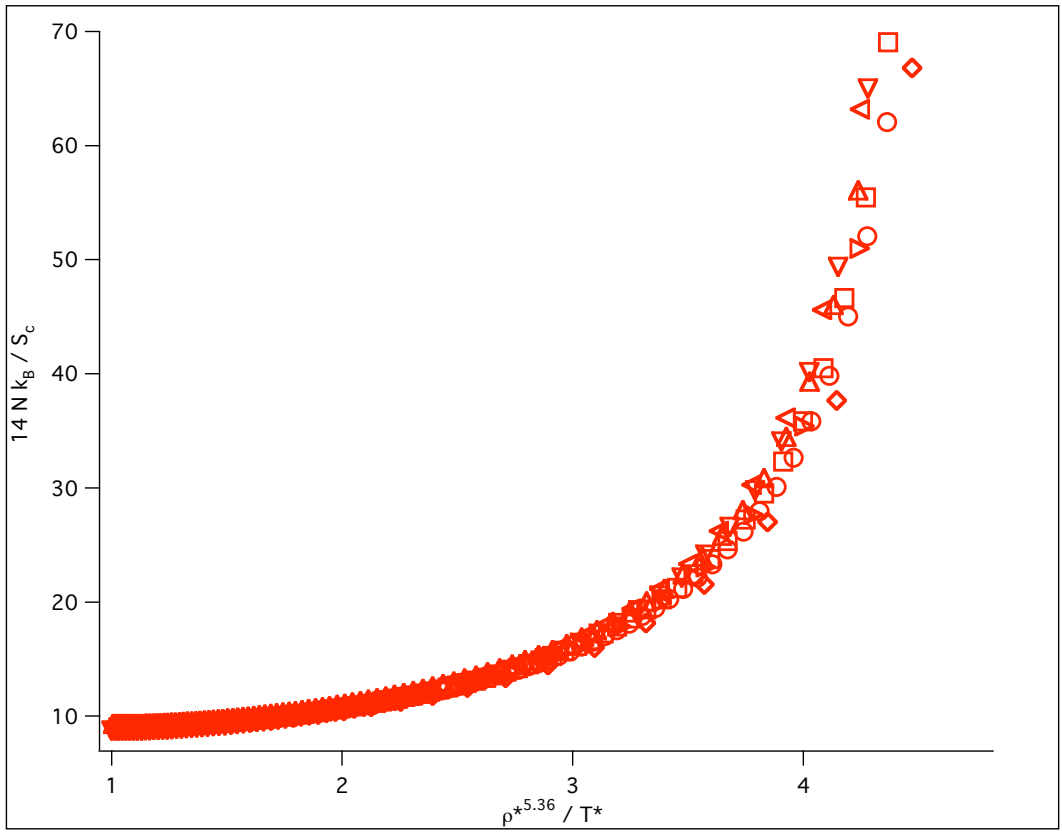}
\caption{Plot of $1/S_{c}$ versus $\rho^{*5.36} /T^{*} $ for $Z_{MGC}$.  The pressures are 25.0 (circles), 20.0 (squares), 15.0 (triangles), 10.0 (inverted triangles), 5.0 (left-facing triangles), 0.0 (right-facing triangles), and -2.0 (diamonds).\label{casalini_pHD1}}
\end{center}
\end{figure}

Figures~\ref{casalini2}-\ref{casalini_pHD1} display plots of $\frac{14 N k_{B} }{S_{c} }$ versus $\rho^{*\gamma}  / T$ using the Mulero, Gal\'{a}n, and Cuadros equation of state.  The values of $\gamma$ shown are best fits to the predicted results. As expected, even though the potential is not a power law the behavior is nearly universal. The predicted values of $\gamma$ do depend on which equation of state is used. For the constant density fits, the analysis yields  similar values for all 3 equations of state ($\gamma = 5.9$ with $Z_{MCG}$, = 5.8 with $Z_{SW}$ , and = 5.9 with $Z_{CS}$).  There is more variance among the values of $\gamma$ when data are examined at  constant pressure. One finds at constant pressure  $\gamma = 5.8$ for the Mulero, Gal\'{a}n, and Cuadros equation of state, $\gamma = 5.5 $ for the Salzburg-Wood equation of state and $\gamma = 4.7$ for the Carnahan-Starling equation of state.  The discrepant latter value is an indication, seen earlier in our calculations of the characteristic temperatures, that the Carnahan-Starling equation of state does not describe the high density glass as well as do the other two equations of state.  This is not surprising, given the well-known deficiencies of this equation of state at high densities.  The range of $\rho^{*\gamma} / T^{*}$ displayed in these figures is much greater than the range accessible to experiment, due to difficulties in obtaining low temperature data.  The parameter $\rho^{*\gamma} / T^{*}$ varies by less than a factor of 2 in experiment; if this lmited range  were used to display our prediction, the appearance of universality would be more striking.  The isochoric values of $\gamma$ are 10\% greater than the corresponding isobaric values of $\gamma$.

\begin{figure}
\begin{center}
\includegraphics[width=6in]{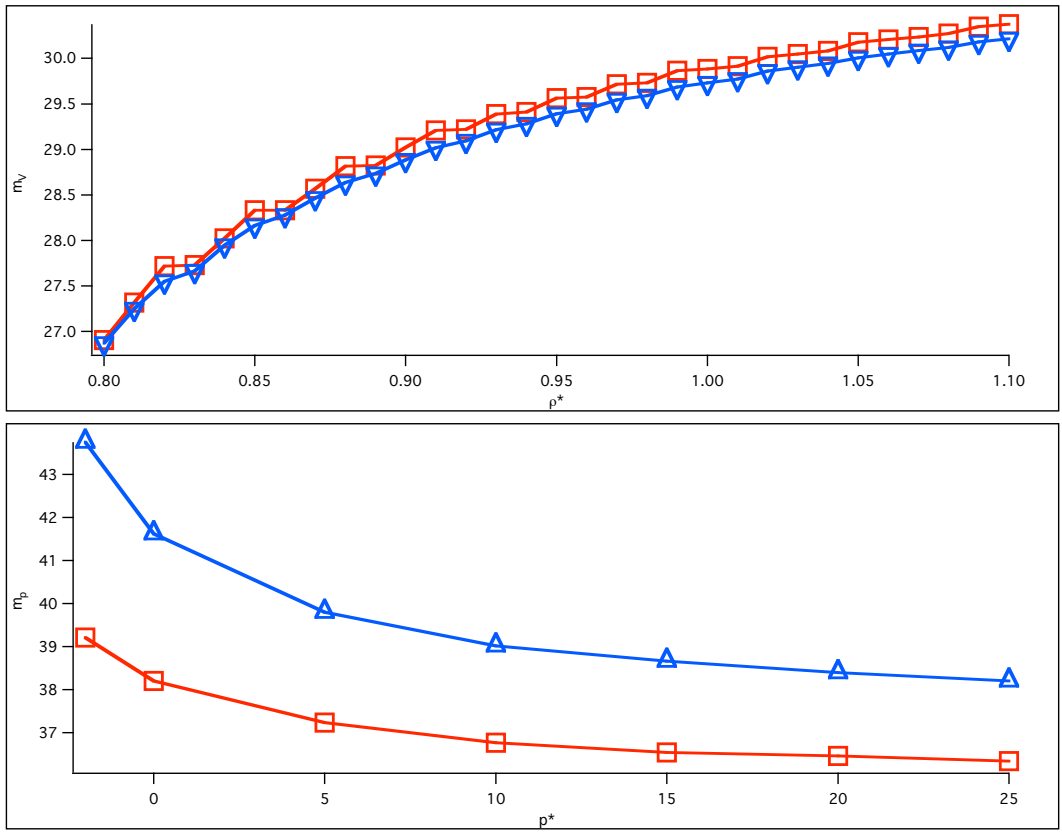}
\caption{Plots $m_{V} $ (top panel) and $m_{p} $ (bottom panel) on the one hour time scale using $Z_{SW}$ (squares) and $Z_{MGC}$ (triangles).\label{mpmv}}
\end{center}
\end{figure}

Casalini and Roland\cite{CasaliniRoland2005} have determined the constant pressure and volume fragilities of several glass formers, with the fragility of a glass being defined as
\begin{eqnarray}
m&=&\left.\frac{d \log_{10}\tau}{d\left(\frac{T_{g} }{T}\right)}\right|_{T_{g} }
\end{eqnarray}
Using the random first order transition theory expressions one finds 
\begin{eqnarray}
m_{V} &=&\frac{14 \Delta C_{V} (T_{g} ) / N k_{B} }{\left(S_{c} (T_{g} ) / N k_{B} \right)^{2} }\\
m_{p} &=&\frac{14 \Delta C_{p} (T_{g} ) / N k_{B} }{\left(S_{c} (T_{g} ) / N k_{B} \right)^{2} }
\end{eqnarray}
As always, the temperature $T_{g}$ to be used in these expressions depends on the time scale of the experiment.  Casalini and Roland evaluate the fragility at a temperature at which the relaxation time is 10 s, instead of the more conventional 1 hour definition of $T^{*}_{g}$.  We have evaluated $m_{V} $ and $m_{p} $ on both time scales; $\Delta f = 0.82$ for the one hour time scale and $\Delta f = 1.069$ for the 10 s time scale.  The one hour time scale results are shown in Figure~\ref{mpmv}. In agreement with experiment (see Fig. 3, Huang, et. al.\cite{McKenna2002} and Fig. 10, Casalini and Roland\cite{CasaliniRoland2005}), there is little density dependence to $m_{V} $, while $m_{p} $ is a decreasing function of pressure.  Direct comparison between theory and laboratory experiment is complicated by small errors in the value of $D$, which we take to be precisely 32, while already the best fit to empricial data would differ by about 10\%. Also one expects differences  between the values of $\Delta C_{p}$ found for specific polyatomic substances and pure Lennard-Jonesium.  In addition, the prediction of absolute fragilities depends quite critically on the corrections made to the free energy of the glassy and liquid configurations.  We note, however, that the ratio $m_{p}/m_{V}$ should be relatively insensitive to all of these effects.  Using $Z_{MGC}$, our calculations give $m_{p=0}/m_{V} \approx 1.5$.  On the 10 second time scale, Casalini and Roland find\cite{CasaliniRoland2005} the ratio equal to 1.9 for salol, 1.4 for propylene carbonate, 2.3 for 1,1'-di(4-methoxy-5-methylphenyl)cyclohexane, 1.9 for phenolphthalein-dimethyl-ether, 1.9 for cresolphthalein-dimethyl-ether, and 2.6 for PCB62, a chlorinated biphenyl. 

\begin{figure}
\begin{center}
\includegraphics[width=6in]{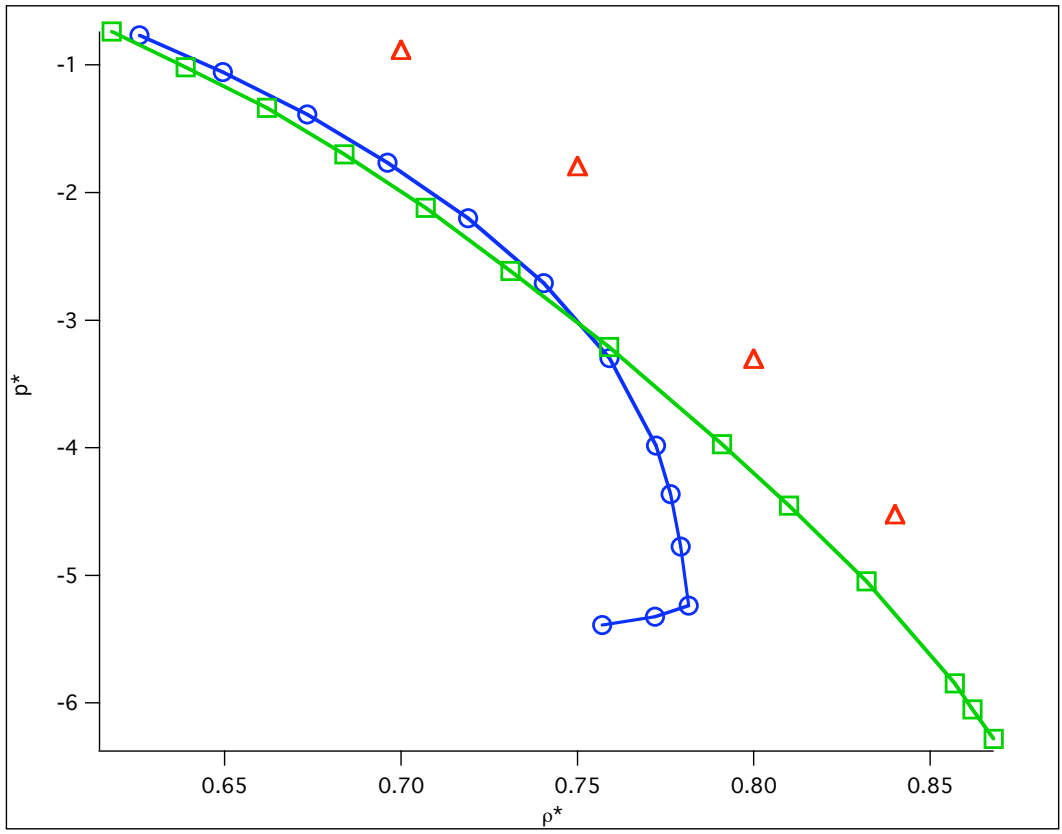}
\caption{Plot of the spinodal pressures versus density, using $Z_{MGC}$ (circles) and $Z_{CS}$ (squares).  Also shown are the spinodal pressures from simulation.\cite{Protsenko2005}\label{pspinodal}}
\end{center}
\end{figure}

\begin{figure}
\begin{center}
\includegraphics[width=6in]{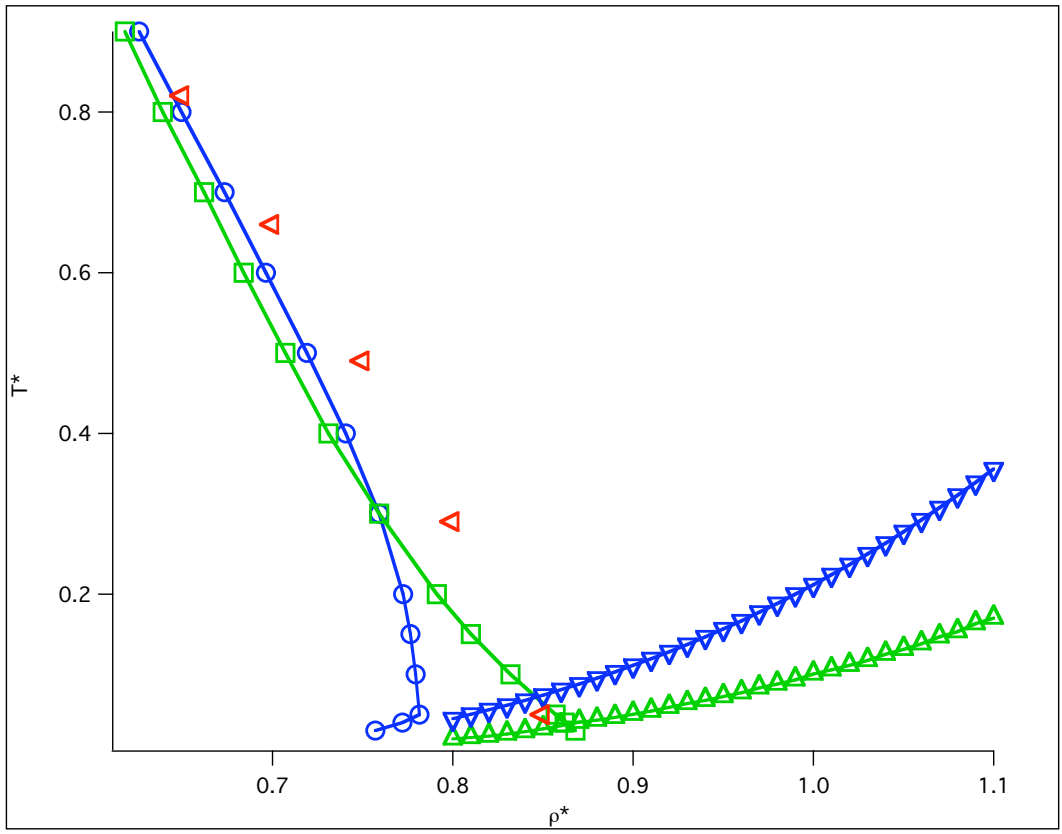}
\caption{Plot of the spinodal (circles, $Z_{MGC}$, and squares, $Z_{CS}$) and Kauzmann (inverted triangles, $Z_{MGC}$, and triangles, $Z_{CS}$) temperatures versus density.  Also shown are the spinodal pressures from simulation.\cite{Protsenko2005}\label{sastryplot}}
\end{center}
\end{figure}

The Sastry crossing of transitions occurs at low densities where we do not expect the $Z_{SW}$ to be accurate.  Therefore our search for the Sastry density employed the equations of state with $Z_{CS}$ and $Z_{MGC}$ only.  We expect the Carnahan-Starling equation of state to be the more reliable one at these lower densities.  $p^{*}_{spinodal}$ and $T^{*} _{spinodal}$ were determined by locating the temperature at which the pressure as a function of temperature was a minimum at a fixed density.  Figure~\ref{pspinodal} displays the spinodal pressure versus density for both equations of state and recent molecular dynamics results of Ba\u{i}dakov and Protsenko\cite{Protsenko2005}.  At these low densities, $Z_{CS}$ agrees well with simulation, while $Z_{MGC}$ deviates from the simulation data above $\rho^{*} \approx 0.75$.  The deviation can be traced to the large value of the blip diameter, $d$, that is found at the low spinodal temperatures using $Z_{MGC}$;  the corresponding values of $\rho^{*} d^{3} $ approach the singularities in $Z_{MGC}$ and render this equation of state unreliable in determining the pressures.
 Figure~\ref{sastryplot} displays the spinodal and Kauzmann temperatures versus density.
 It is clear that both equations of state predict a crossing of the spinodal and Kauzmann lines, at low temperatures, but disagree moderately as to  the density of the crossing. $Z_{CS}$ predicts a crossing at a density of $\approx$ 0.85 while $Z_{MGC}$ predicts a crossing at $\approx$ 0.80.  Both values are in good agreement with the estimates from simulation\cite{Protsenko2005}.

\section{Conclusions}
Random first order transition theory as instantiated microscopically with  self-consistent  phonon theory and perturbative liquid state theory has been used to predict characteristics of the glassy dynamics of a supercoooled Lennard-Jones liquid. The calculations are seen to be largely robust with respect to the choice of liquid equation of state and are in good agreement with inferences from simulation and experiment. Isochoric and isobaric calculations have been performed to distinguish those quantities arising  from purely steric interations from those due to attractive interactions.  We find that the trends of configurational heat capacities are reproduced well, but the overall magnitudes are quantitatively 10-20\% too small.  The calculated fragilities follow experimental trends for the density and pressure dependencies (demonstrating the present microscopic calculations can accurately reproduce the sign of the third derivative of the configurational contributions to the free energy!).   Further, the spinodal line, as a function of density, is found to intersect the Kauzmann temperature line at low densities and temperatures, in agreement with the speculation of Sastry. 

Random first order transition theory is thus able to explain a variety of interrelated features of glasses that are directly related to the intermolecular forces, including explaining the density-temperature scalings highlighted by Casalini and Roland and the low density behavior described by Sastry.  The small, but not absent, dependence of the fragility on density seen in experiment is reproduced by the theory and the ratio of the isobaric to isochoric fragilities is also in reasonably good agreement with experiment.

The strategy we employ here also provides a route to quantify glassy characteristics of complex polyatomic molecular fluids.  Molecular liquid theory provides expressions for the equations of state for polyatomic fluids that can be combined with self-consistent phonon theory to extend the present treatment to many of the specific molecular liquids commonly studied in the laboratory.

\section*{Acknowledgment.}
The authors wish to acknowledge the support of NSF grants CHE-0317017 and CHE-0517236(RWH).  Calculations were performed on computational facilities provided by LSU (http://www.hpc.lsu.edu) and the Louisiana Optical Network Initiative (http://www.loni.org).

\section*{References and Notes}

\newpage
\section{Tables}

\begin{table}[htbp]
\caption{$\Delta C_{V} $ at $T_{g} $ for the three equations of state used.}
\label{cv_table}
\begin{tabular}{|l|c|c|c|c|c|c|}
\hline
&\multicolumn{2}{|c|}{$Z_{SW}$}&\multicolumn{2}{|c|}{$Z_{MGC}$}&\multicolumn{2}{|c|}{$Z_{CS}$}\\ \hline
$\rho^{*}$&$T^{*}_{g} $&$\Delta C_{V} / N k_{B} $&$T^{*}_{g} $&$\Delta C_{V} / N k_{B} $&$T^{*}_{g} $&$\Delta C_{V} / N k_{B} $\\ \hline
1.10&0.53&1.47&0.54&1.46&0.38&0.92\\ \hline
1.05&0.42&1.46&0.43&1.45&0.29&0.92\\ \hline
1.00&0.32&1.45&0.33&1.44&0.22&0.90\\ \hline
0.95&0.24&1.43&0.24&1.42&0.16&0.91\\ \hline
0.90&0.17&1.40&0.17&1.40&0.12&0.89\\ \hline
0.85&0.11&1.37&0.12&1.36&0.08&0.88\\ \hline
0.80&0.07&1.30&0.07&1.30&0.05&0.78\\ \hline
\end{tabular}
\end{table}

\begin{table}[htbp]
\caption{$\Delta C_{p} $ at $T_{g} $ using $Z_{SW}$.}
\label{cp_table}
\begin{tabular}{|l|c|c|c|c|c|c|}
\hline
&\multicolumn{2}{|c|}{$Z_{SW}$}&\multicolumn{2}{|c|}{$Z_{MGC}$}&\multicolumn{2}{|c|}{$Z_{CS}$}\\ \hline
$p^{*} $&$T^{*}_{g} $&$\Delta C_{p} / N k_{B} $&$T^{*}_{g} $&$\Delta C_{p} / N k_{B} $&$T^{*}_{g} $&$\Delta C_{p} / N k_{B} $\\ \hline
25.0 &0.77&1.76&0.77&1.85&0.64&1.22\\ \hline
20.0 &0.68&1.76&0.69&1.86&0.57&1.24\\ \hline
15.0 &0.59&1.77&0.59&1.87&0.50&1.25\\ \hline
10.0 &0.49&1.78&0.50&1.89&0.42&1.27\\ \hline
 5.0 &0.39&1.80&0.39&1.93&0.33&1.31\\ \hline
 0.0&0.26&1.85&0.27&2.01&0.24&1.41\\ \hline
 -2.0&0.20 & 1.89&0.21&2.12&0.19&1.52 \\ \hline
 \end{tabular}
 \end{table}

\newpage
\section{Figures}

\end{document}